\documentclass[twocolumn]{emulateapj}
\usepackage{amsmath}
\usepackage{wasysym}
\usepackage{pstricks}

\slugcomment{First Submitted on 18/12/2012, Accepted for Publication in  APJ on 11/09/2013}

\shorttitle{Large-scale Magnetic Structure Formation}
\shortauthors{Malapaka et al.}

\begin{document}


\title{ Large-scale Magnetic Structure Formation in 3D-MHD Turbulence}


\author{Shiva Kumar. Malapaka\altaffilmark{1,2} and Wolf-Christian M\"uller\altaffilmark{1,3}}
\affil{1: Max-Planck Institute for Plasmaphysics, Boltzmannstrasse 2, D-85748, Garching bei Muenchen, Germany. \\
 2: Department of Applied Mathematics, School of Mathematics, Univeristy of Leeds, Leeds, LS2 9JT, UK.\\
 3. Zentrum f\"ur Astronomie und Astrophysik, TU-Berlin, ER 3-2, Hardenbergstr. 36, 10623 Berlin.}




\begin{abstract}
The inverse cascade of magnetic helicity in 3D-MHD turbulence is believed to be one of the processes responsible for large scale magnetic structure formation in astrophysical systems. In this work we present an exhaustive set of high resolution direct numerical simulations (DNS) of both forced and decaying 3D-MHD turbulence, to understand this structure formation process. It is first shown that an inverse cascade of  magnetic helicity in small-scale driven turbulence does not necessarily generate coherent large-scale magnetic structures. The observed large-scale magnetic field, in this case, is severely perturbed by magnetic fluctuations generated by the small-scale forcing. In the decaying case, coherent large-scale structure form similar to those observed astronomically. Based on the numerical results the formation of large-scale magnetic structures in some astrophysical systems, is suggested to be the consequence of an initial forcing which imparts the necessary turbulent energy into the system, which, after the forcing shuts off, decays to form the large-scale structures. This idea is supported by representative examples e.g. cluster of galaxies. 
\end{abstract}





\section{Introduction}
Large-scale magnetic structures are observed in many astrophysical systems, for example in galaxies, galaxy clusters, or around stars and planets \citep{sce07,bag09, Bel09}. The scales of these structures can range from megaparsecs on the farther side to few astronomical units on the smaller side. A common feature of these systems is plasma turbulence. Theories explaining the formation of these structures use the concepts of magnetohydrodynamics (MHD) and MHD-turbulence (see \cite{bis03,Bransub05}). Prominent theories include the dynamo effect to explain the magnetic fields of planets, stars and galaxies (see e.g.\cite{cow81,san99,wei02,Bra02,zh04,sub06,sol06}) and feeding of plasma through jets in the case of clusters of galaxies \citep{con09}. However, no single theory or phenomenology is capable of explaining the formation of these structures, consistently. \\
 Although these theories succeed in explaining the amplification of a seed field into large-scales, the stability of these large-scale magnetic structures in a highly turbulent environment (e.g. see images of MRC 0116+111 in \cite{bag09}), remains an issue that is less understood \citep{wei02, bis03}.\\
It is envisaged that the inverse cascade of magnetic helicity $H^M$ in three-dimensional MHD turbulence, i.e. spectral transfer of $H^M$ from small scales to large scales, with a constant flux, is one of the mechanisms that can explain large-scale magnetic structure formation and their stability \citep{Po76}. Magnetic helicity is defined as $\frac{1}{2} \int dV \mathbf{a} \cdot \mathbf{b}$, where $V$ is the overall volume, $\mathbf{a}$ is the magnetic vector potential and $\mathbf{b}$ is the magnetic field. Although there have been several works discussing the formation of large-scale structures using the inverse cascade of magnetic helicity as one of the mechanisms (see e.g \cite{gil89,kin95, mil96, fie00, al07}), only a few studies have given this concept the `central stage' (see e.g. \citep{Bra01, min06, min07}). However, even in these numerical studies, the energy-containing scales of the flow are not separated enough from the scale of the system, thus not allowing us to get a clear picture of the unique properties of a self-similar inverse cascade --- in short, there is not enough scale separation in these numerical studies. In contrast, in many of the astrophysical systems, the scales characteristic of magnetic structures are orders of magnitude larger than the energy containing scales of the associated turbulent plasma flow (for example: jets in the case of  radio relics and cores of the planets in the case of planetary magnetospheres).\\
Hence, the primary objective of this work, is to study the inverse cascade of magnetic helicity as a possible fundamental formation process of large-scale magnetic structures using direct numerical simulations. For this, we use an idealized system of plasma flow comprising of statistically isotropic, homogeneous, incompressible three dimensional MHD turbulence with vanishing mean alignment of $\mathbf{v}$ (the velocity field) and $\mathbf{b}$. The scale separation between the turbulence drive and the system size is made as large as the numerical resources permit, by using a periodic cubic grid of $1024^3$.\\
This paper is divided into 6 sections. In section 2 the model equations used and numerical set up are described, followed by a section summarizing pertinent results of spectral studies. The next section discusses obtained numerical results based on data visualization for driven and decaying MHD turbulence drawing conclusions on a possible fundamental process of large-scale structure formation. In section five an attempt is made to explain magnetic structure formation in some astrophysical example systems. Conclusions are given in section six. 
\section{Numerical set up}
The set of dimensionless incompressible MHD equations giving a concise single-fluid description of a plasma and the numerical setup used to solve them have already been discussed in our works on spectral studies \citep{mul12,mulmal12} and are repeated below for completeness as:
\begin{subequations}
\label{solcon}
 \begin{equation}
 \partial_{t} \boldsymbol{\omega} =
 {\boldsymbol \nabla}\times({\bf v}\times{\boldsymbol\omega} - {\bf b}\times{\bf j})+
 {\mu}_{{n}}(-1)^{n/2-1}\nabla^{{n}}
 {\boldsymbol\omega}+{\bf {F_{v}}}+{\lambda}\Delta^{-{1}}{\boldsymbol\omega}\,,
 \end{equation}
 \begin{equation}
\partial_{t}{\bf b} = {\boldsymbol \nabla}\times({\bf v} \times {\bf b}) +
 {\eta}_{{n}}(-1)^{n/2-1}\nabla^{{n}}
 {\bf b}+{\bf {F_{b}}}+{\lambda}\Delta^{-{1}}{\bf b}\,,
\end{equation}
\begin{equation}
 {\bf {\nabla\cdot v}} = {\bf {\nabla\cdot b}}= {0}
\end{equation}
\end{subequations}
where $\boldsymbol{\omega} = \nabla \times \bf{v}$ is the vorticity and $\bf{j} = \nabla \times \bf{b}$ is the current density.\\ 
 Relativistic effects are neglected and the mass density is
 assumed to be unity throughout the system. Other effects
 such as convection, radiation and rotation are also neglected.
 Direct numerical simulations are performed by solving the 
 set of model equations by a standard pseudospectral method \citep{can88}
 in combination with leap-frog integration 
 on a cubic box of linear size $2\pi$ that is discretized with $1024$ collocation points 
 in each spatial dimension. 
 Spherical mode truncation is used for alleviating aliasing errors. 
 By solving the equations in Fourier space, the solenoidality of ${\bf v}$ and ${\bf b}$
 is maintained algebraically. \\
 As the primary objective of our study is to observe clear signatures of an inverse cascade of magnetic helicity 
a source of this quantity is introduced at small scales. This is achieved in two different ways resulting in two main configurations:
a driven system and a decaying one. In the driven case, the forcing terms 
${\bf {F_v}}$ and ${\bf {F_b}}$ are delta-correlated random processes 
acting in a band of wavenumbers $203\leq k_0\leq 209$. They create a 
small-scale background of fluctuations with adjustable amount of magnetic and 
kinetic helicity ($H^{K} =\frac{1}{2}\int_V{\bf {v\cdot \boldsymbol{\omega}}}\,{\mathrm d}V$). The results reported in this paper do not change if kinetic helicity injection is finite. We also report the results from a third configuration wherein the system is initially forced and later allowed to decay.
The theoretical results presented in the following do not depend on 
the setup of the forcing as they presuppose an existing self-similar distribution of
energies and helicities. For obtaining such spectra in numerical experiments 
the magnetic source term ${\bf {F_b}}$ is necessary while a finite momentum source ${\bf {F_v}}$ 
speeds up the spectral development significantly.
In the decaying case the forcing terms are set to zero and the initial condition represents an
ensemble of smooth and random fluctuations of maximum magnetic helicity with respect to the energy
content (for details see \cite{mulmal12}) and a characteristic wavenumber $k_0=70$.  The third configuration is obtained by withdrawing the forcing at (few) selected points in the evolution of the driven system and advancing the obtained configurations as decaying systems, for (at least) further five large eddy turnover times .\\
 A large scale energy sink $\lambda\Delta^{-1}$ with $\lambda=0.5$ is present for both the fields in the forced case, to ensure the minimization of finite volume effects on the inverse cascade. In the decaying case $\lambda$ is set to zero and hence the modifications due to finite system-size on the scaling regime of the cascade are reduced by stopping the simulation(s) as the largest-scales of the box are reached. The simulations are run for $6.7$ (forced) and $9.2$ (decaying) large-eddy turnover times of the system respectively, with the time unit being defined using the system size and its total energy. The hyperdiffusivities ${\mu}_{{n}}$ and ${\eta}_{{n}}$ are dimensionless dissipation coefficients of order $n$ , which is always even in these simulations), with $n=8$ in both runs. 
They act like higher-order realizations of viscosity and magnetic diffusivity, respectively. 
The magnetic hyperdiffusive Prandtl number ${Pr_m}_n={\mu_n}/{\eta_n}$ is set to unity.\\
The initial conditions to these simulations are smooth fluctuations with random phases having a Gaussian
energy distribution  peaked around $k_0$ in the decaying and the forced cases. In the present driven-simulations delta-correlated random processes generate
magnetic and velocity field fluctuations with wavenumbers $203<k_0<209$ and well defined kinetic and 
magnetic helicity. In all reported simulations magnetic driving with maximal magnetic helicity of one sign 
is applied while the velocity driving does not inject kinetic helicity.
Test runs with additional kinetic helicity injection (various levels relative to 
magnetic helicity injection niveau) have been carried out without significantly changing 
the described observations.
Control of the respective  helicity input is achieved by constructing the injected 
fluctuations as linear combinations of eigenvectors of the curl operator.
Technical details are given in \cite{Bis00} and \cite{wa192}.
The initial/force-supplied ratio of kinetic to magnetic energy is unity with a peak amplitude of 0.05 in the forced
 case and an amplitude of unity in the decaying case. In the third case, this ratio depends on the selected point in the evolution of the driven system. Hyperviscosity of order $n=8$ is chosen in the simulations to obtain sufficient scale-separation. It is difficult to define an unambiguous Reynolds number owing to a) the scale separation and b) the use of hyperviscosity in the system, as several different Reynolds numbers could be defined (estimated) at several different characteristic scales (also see for more details \cite{Mal09}). For example, if we presume that the characteristic length scale $\it{l}$ of the system is defined in the inertial range for both forced and decaying cases, then the Reynolds number could be defined as $\it{Re^{\frac{9}{4}}}$=$\frac{\it{l}}{\it{l_d}}$ where $\it{l}$ equivalent to the characteristic length scale obtained using $(\frac{{\mu}_{n}^{3}}{\epsilon})^{\frac{1}{6n-2}}$ (see \cite{bis03})), with $\epsilon$ being the constant non-linear energy flux and $\it{l_{d}}$=$(\frac{{\eta}_{n}^{2} \it{v_{A}}}{\epsilon})^{\frac{1}{4n-1}}$ being the dissipation length with $v_A$ the Alfv\'en velocity calculated from $\sqrt{E^M}$ at the characteristic length scale $\it{l}$. The estimated Reynolds number in this method is in the range of  $10^3$ to $4\times 10^3$ in all the three cases.\\
 With the above mentioned simulation set up, the equations are solved for all the three cases separately in the spectral space and using inverse Fourier transforms, spatial data is obtained. With the help of this data, we discuss the evolution of large-scale magnetic structures and interpret them for various astrophysical scenarios, in the following sections.
\section{Spectral behavior of magnetic helicity and magnetic energy}
 Magnetic helicity is an ideal invariant in 3D-MHD which via its spectral behavior helps in gaining some insight into the turbulent MHD flow. It quantifies the linkage and twist in the magnetic field \citep{Mof69}. The change of magnetic field topology associated with a spectral transfer of magnetic helicity can be caused by spectrally local or non-local nonlinear dynamics. \cite{al06} report that the cascade of magnetic helicity in direct numerical simulations of forced 3D-MHD turbulence, has both local and non-local components and that the inverse cascade process is predominantly non-local in nature. In \cite{mul12}, it is shown that the inverse cascade of magnetic helicity in the forced case has a large scale  self-similar range ($\it{k} \sim 7 - 30$) with a power law exponent of -3.3, while it is -3.6 for the decaying case \footnote{In the decaying case, the spectral transfer of magnetic helicity to smaller $\it{k}$ or large scales, cannot be termed as a cascade as the flux of helicity in this case is not a constant. It can only be termed as inverse spectral transfer(see \cite{Mof78,chr05,Mal09}). However, the use of "inverse spectral transfer" might confuse the readers and to avoid this `inverse spectral transfer' is used sparingly and the word `inverse cascade' is used in most of the contexts.}. In the small scale region ($\it{k} \sim 250 - 400$) we observe a second self-similar range with a power law of -1.7, arising out of a direct cascade of magnetic helicity. This small-scale self-similar range needs to be treated with caution as the spectra in this range might be influenced by bottleneck effect \citep{Bis00}, and hence we will ignore the discussion on this range in the following section. Our results on the power law of magnetic helicity differ significantly from the previous DNS attempts (see for example \citep{Bra01, min06}) which did not have a clear scale separation between the large and small scales. However, recently there have been several results in the decaying case e.g. \citep{min10, john11}, which confirm the deviation of the power law value of magnetic helicity from the earlier predicted value of $\it{k}^{-2}$. The power law values obtained in these new simulations range from $\frac{-10}{3}$ to $-3.7$, suggesting a lack of universality for the power law of magnetic helicity \citep{pouq10}. Owing to the range this new power law falls in all these works [-3.3 to -3.7] and the fact that even to the present day, the best resolved numerical simulations still permit an unphysical dependence of the system behavior on initial conditions and a specific simulation setup, we observe that it is probably too early to comment on the lack of universality, or even the true value of the scaling exponent. One unambiguous comment that can be made here is that the physics behind the self-similar scaling exponents is not clearly understood, as the earlier dimensional analysis argument using magnetic helicity flux (see \cite{Po76}) appears to be no longer valid in these works.\\
 The inverse cascade of magnetic helicity with periodic boundary conditions has been studied in the context of mean field dynamo theory and $\alpha$-effect. There, it was realized that $\alpha$-effect gets quenched due to the presence of periodic boundaries in combination with magnetic helicity conservation and to overcome this, simulations with open boundaries; to shed the helicity flux from the system; were proposed (see \cite{Bra04,Bransub05}). However, we in this paper are not dealing with the constraints imposed by the cascade of magnetic helicity on the dynamo process. Our main focus here is to understand the inverse cascade as a `transport process', moving the small-scale magnetic helicity flux non-locally, aiding in the generation of large scales.\ 
The inverse cascade of magnetic helicity transports several other quantities; which show predominantly local non-linear interactions e.g. magnetic energy, kinetic energy (as a result total energy) and kinetic helicity, along with it to larger scales (see \cite{mul12, mulmal12} and the references therein).
 As the objective is to understand the formation of large-scale magnetic structures, the influence of the inverse cascade of magnetic helicity on the magnetic energy spectrum is described below.
\subsection{ Magnetic Energy Spectrum}
\begin{figure}
\begin{center}
\includegraphics[width=0.85\textwidth,viewport=100 115 530 760,clip]{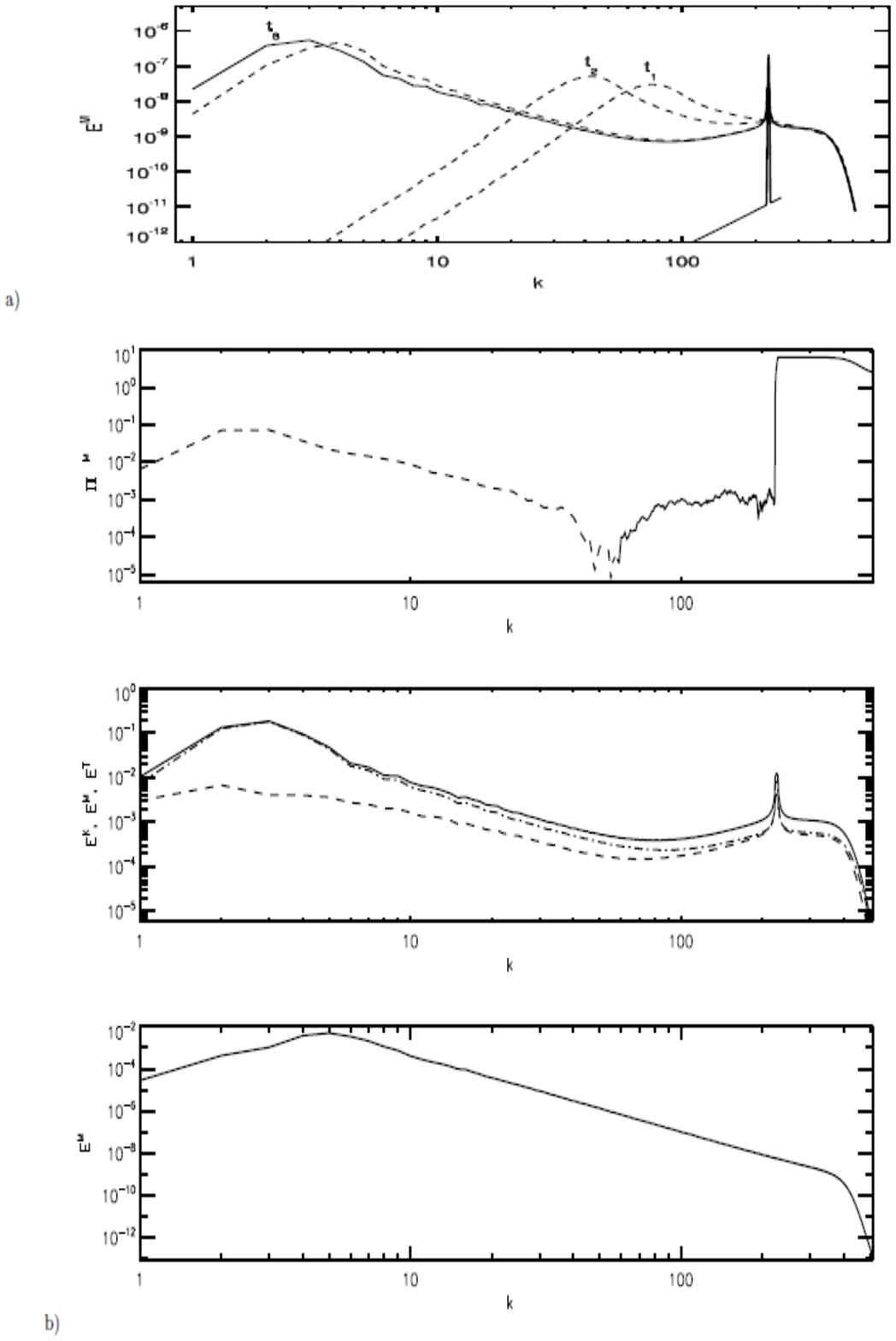}  
\caption{ Spectral properties of magnetic energy and comparison of energies. a) Spectra of magnetic energy in the forced case plotted at four instances of time. Three instances as ${t_1} = 0.143$, ${t_2} = 0.33$ and ${t_3}=6.7 $ are marked on the plot(see sections 3 and 4.4), b) flux of magnetic energy plotted at $t_3$ in the forced case (top) solid line: positive flux and dashed line: negative flux, comparison of energies at $t_3$ in the forced case (middle). Kinetic energy $E^K$ (dashed line), magnetic energy $E^M$ (dash-dot line) and total energy $E^T$ (solid line) and magnetic energy spectrum in the decaying case at $t=9.2$ (bottom). A similar trend for both the plots (top and middle) is seen in the decaying case, hence they are not shown here.}
\label{fig1}
\end{center}
\end{figure}

Magnetic energy and magnetic helicity are dimensionally related as $\it{E}^{M}_{\it{k}}\sim \it{k}\it{H}^{M}_{\it{k}}$. It is indeed observed that as the magnetic helicity inverse cascades, so does the magnetic energy (shown in Figure (1a)), but with a less steeper (approximate) power law of $\it{k}^{-1}$ \citep{Po76}. In this work too, we observe approximately the same relation between magnetic energy and magnetic helicity with magnetic energy showing a large scale $(\it{k} \sim 7 - 30)$ self-similar range with the power-law scaling exponent having a value of -2.1, in the driven case. In the small-scales ($\it{k} \sim 250 - 400)$, the scaling exponent's value is -0.6. In the decaying case the self-similar range has a power law scaling exponent of -2.1 (Figure (1b bottom)). It is to be noted that the flux of magnetic energy; $\Pi_{jb + vb}(\it{k})= \int^{\it{k}}_{0}{\it{d^{3}}\it{k'}[\underbrace{(\bf{Re}({\tilde{\boldsymbol{\omega}}^{*}.\frac{\it{i}}{\it{k'^{2}}}(\it{k'}\times\widetilde{\overline{\bf{j}\times\bf{b}}})}))}_{T_{jb}}+\underbrace{(\bf{Re}(\tilde{\bf{b}}^{*}.\it{i}(\it{k'}\times\widetilde{\overline{\bf{v}\times\bf{b}}})))}_{T_{vb}}]}$
 where $T_{jb}$ is the contribution of the magnetic field to the flux and $T_{vb}$ is a contribution to the flux due to the interaction of both velocity and magnetic fields (here the terms with $\sim$ are the Fourier transformed quantities),is not constant and depends on wave number $\it{k}$ as shown in Figure (1b top).\\
It has been observed that although the simulation in both decaying as well as forced cases start with the same initial kinetic and magnetic energies, as the system evolves in time, the magnetic energy gains at the cost of kinetic energy at large scales, a behavior known and understood as a result of a dynamical equilibrium of small-scale dynamo action and energetic equipartition through Alfv\'en waves \citep{mulgra} (also see Figure (1b middle)). Further, there exists a dynamic relation combining magnetic helicity and magnetic energy with kinetic helicity and kinetic energy, in the spectral space, that shows additional interesting similarities to the to the $\alpha$ dynamo and its quenching \citep{mul12, mulmal12}.\\
 In the third configuration where the system was initially forced and later allowed to decay, the spectral powers changed from the forced configuration values to the decaying configuration values.\\
From the discussion above it can be argued that large-scale magnetic structure formation can occur in both forced and decaying cases, as both the magnetic helicity spectrum and the magnetic energy spectrum move towards the large scales with time. However, only structure function analysis and (or) visualization of the real space structures can give us a true picture. Structure function analysis and other statistical methods are reported else where (see \cite{mal13}) and here in this report we discuss only the results from the visualization of the real space structures from our simulations. However, an analysis of the PDFs of the observed magnetic structures is presented to understand the nature of coherence in these structures.
\placefigure{fig1}
\section{Visualization of the Structures}
 The real-space structures of the simulation data is visualized using tools like AMIRA now known by the name AVIZO (a commercial visualization package) and Visit (an open source visualization package developed by LLNL, USA), to understand the various types of structures present in the turbulent flow. Below, we present a brief summary of systematic study on the observed structures in all the three cases mentioned above.\\
 
 \subsection{Structures in Forced turbulence}
\begin{figure}
\begin{center}
\includegraphics[width=0.9\textwidth,viewport=55 275 538 715,clip]{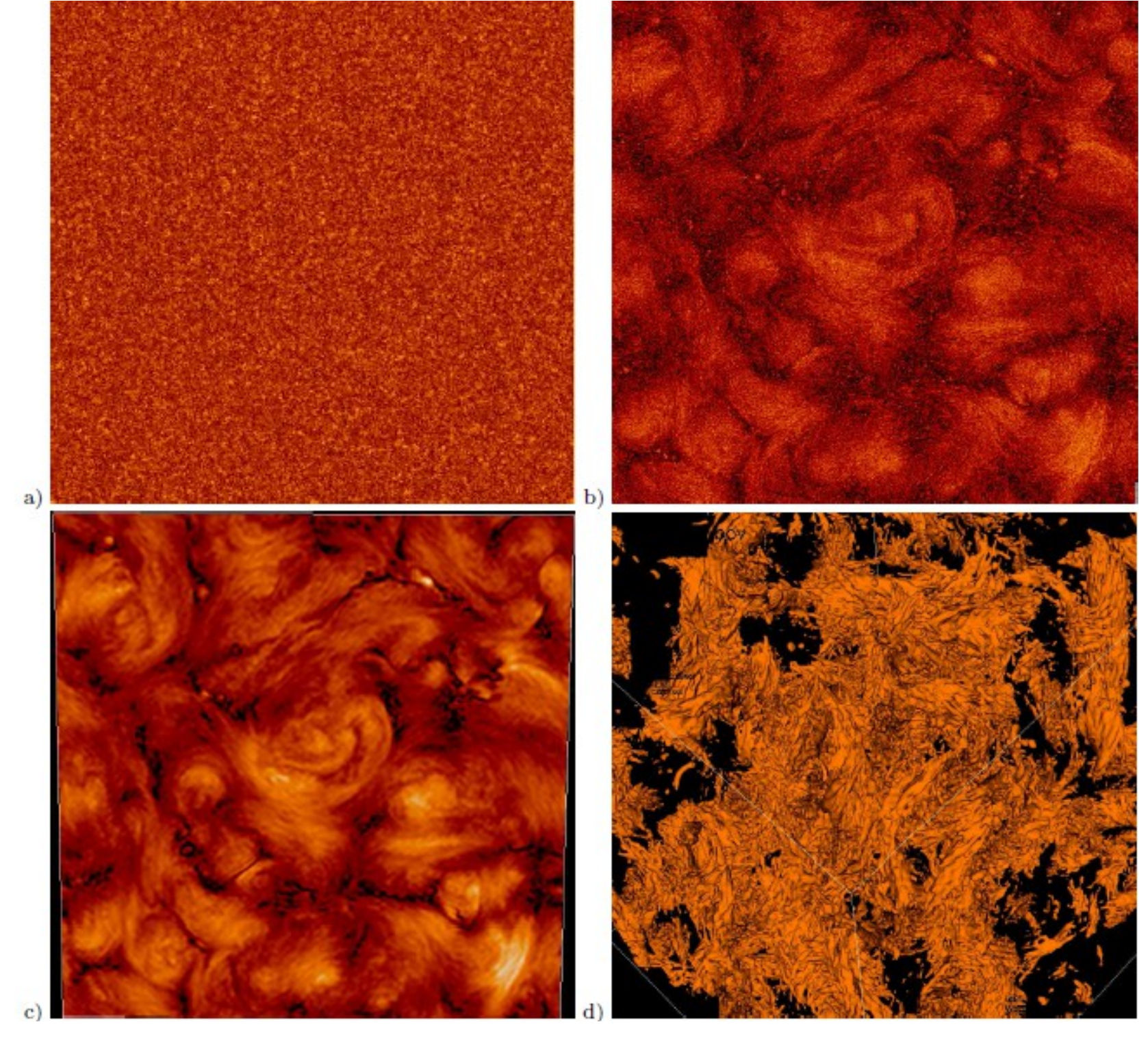}
\caption{Magnetic field structures and iso-surfaces in the forced case. a) Cut across the plane of initial state of unfiltered magnetic field structures b) final state of the unfiltered magnetic field structures  at $t_3$ c) cut across the plane of cut-off filtered out put of magnetic field structures and d) iso-surfaces of the cut-off filtered out put of magnetic field. All the plotted values are absolute values of the fields.  (A color version of this figure is available in the online journal.)}
\label{fig2}
\end{center}
\end{figure}

 The magnetic field structures\footnote{All the structures shown here are color-coded modulus of the respective field and iso-surfaces are are taken at a different threshold value for each field. The brighter regions in all the figures infer higher values and darker regions indicate lower values.} in the forced turbulence case are visualized at $t=0$, the initial state and at a time $t=6.7$, when the peak of magnetic helicity spectrum, has reached the farthest scale (i.e. $\it{k} =2$). The initial state as can be observed from the Figure (2a), consists of small-scale random structures. From the  spectral analysis, it is seen that starting with this initial state, as simulations progress in time, a movement of magnetic energy towards the larger scales along with the inverse cascade of magnetic helicity. Such a spectral transport is expected to aid in the development of coherent large-scale magnetic structures \citep{Po76}.
\placefigure{fig2}
However, it appears that although there is large-scale structure formation (Figure (2b)), these structures are not the usually expected coherent large-scale structures. The probable cause for this unexpectedly looking structures could be the influence of the small-scale random, but energetic forcing supplied at every time step, which seems to appear as some form of dominant noise in Figuure (2b). To remove the reminiscence of the dominant small-scale seeds generated by the forcing and to examine if there are any coherent large scale features hidden behind, the data is subjected to a low pass filtering ($\it{k_{cutoff}} \sim$ 70). The filter is characterized as follows:\\
\begin {equation*}
{
\it{f}(\it{k}) = \begin{cases}\it{f(k)} & $\hfil$  \text{if} $\hfil$ \it{k}\leq 70 \\
 0,$\hfil$ & \text{otherwise.}
 \end{cases}}
\end {equation*}\\
The filtered output devoid of the small scales, is shown in the Figure (2c) and iso-surfaces of the same are seen in the Figure (2d). From these two pictures it can be inferred that the magnetic field has several smudged regions of high concentration which are not the expected coherent large-scale structures. The iso-surfaces of filtered data shown in Figure (2d). indicate that no clearly discernible large-scale magnetic structures have formed up to this point in time. Intermittency modeling of the iso-surfaces (see \cite{Mal09}) concludes that they are fractal in nature (having a co-dimension of 1.5). Thus it appears that, inverse cascade of magnetic helicity does tend to produce large-scale magnetic structures, but the inherent small-scale fluctuations of the forcing imposes its randomness on the self-organization process of the magnetic field. Previous studies (e.g. \cite{Bra02}) have shown that inverse cascade of magnetic helicity in the driven case is supposed to develop over resistive time scales (say $t_R$). Thus, it seems plausible that in this case, evolution of coherent large-scale structures could be of the order of resistive time scales, over which the random affects of the forcing are also effectively overcome. But to perform such a numerical simulation, one would require much larger simulation grids and simulation times; to allow for further development of scales; than the ones that evolved in the presented case. The applied large-scale energy sink necessary to prevent condensation of magnetic helicity on largest system scales also represents an artificial influence whose is impact is difficult to estimate. However, even with the above mentioned uncertainties in the performed simulations, the physical conclusions reached here, represent an important advancement in the understanding of the driven MHD flows as scale-to-scale development is observed at a hitherto unprecedented detail.\\
 In contrast to the initial conditions and driving placed in the small-scales in this work, \cite{Bra01} uses initial conditions and forcing mechanism placed in the `large-scales' that develop into further (coherent)large-scales, although there is not enough scale separation for a sufficient inverse cascade of magnetic helicity to develop,in the spectral space. Thus, the nature of the formed structures in the forced MHD turbulence appears to be severely depend on the type of initial conditions and forcing mechanism (either large-scale or small-scale) used in the simulations. To get further insights on this issue, it is important to study forced MHD-turbulence, with several other types of initial conditions and forcing mechanisms, placed at various different scales.\\
 Other physical quantities like velocity, current density and vorticity were also visualized. Their spatial structure resembles that of the magnetic field and hence are not presented here. Magnetic helicity, however shows large-scale structures which are very similar to the ones observed in the decaying case and hence will be discussed in the section below on the decaying case.
\subsection{Decaying Case}
\begin{figure}\nonumber
\begin{center}
\includegraphics[width=0.9\textwidth,viewport=50 115 535 775,clip]{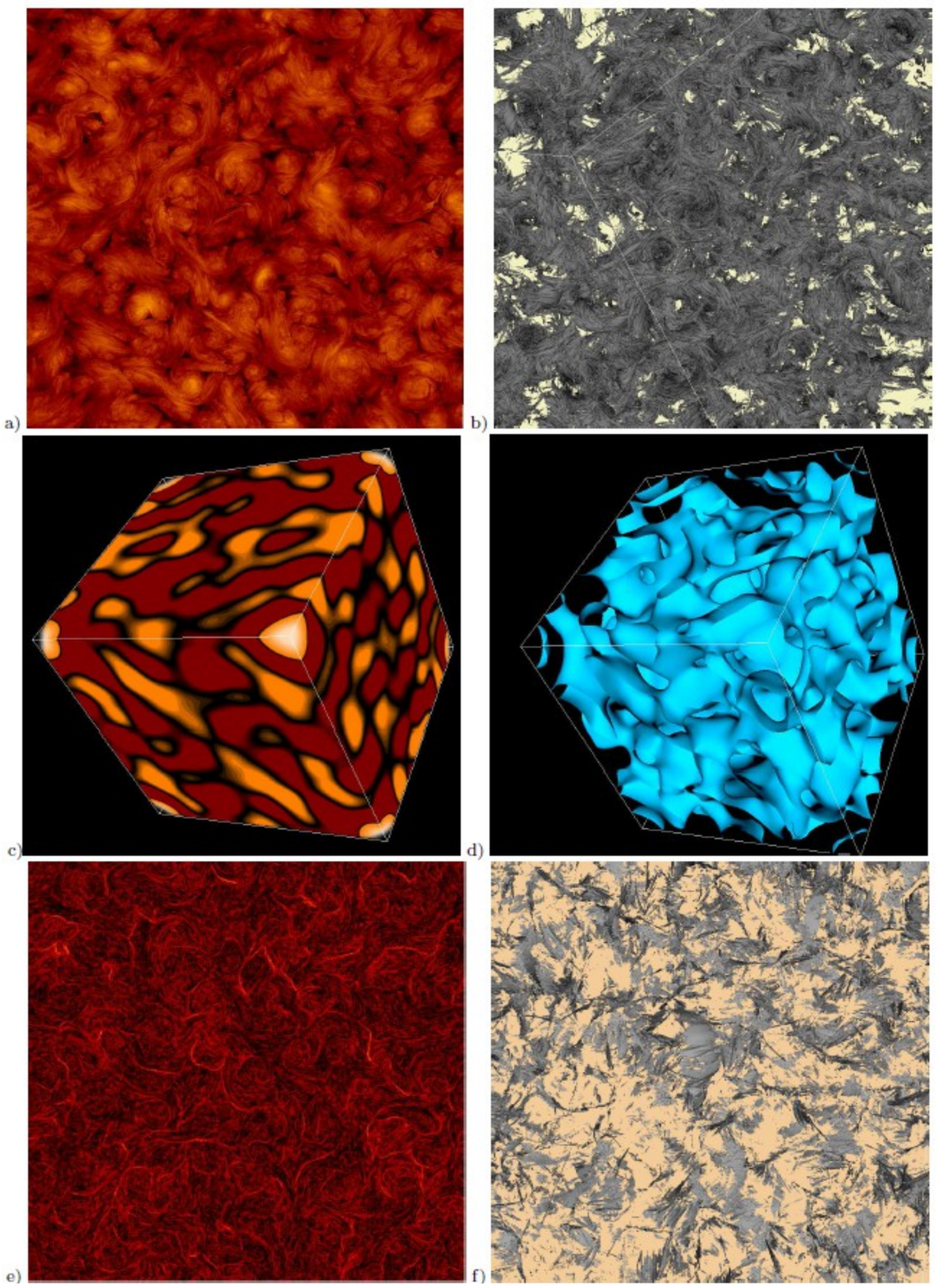}\\
 \scriptsize{Figure caption in the next page} 
 \label{fig3}
\end{center}
\end{figure}

\begin{figure}
\begin{center}
\includegraphics[width=0.9\textwidth,viewport=60 320 540 765,clip]{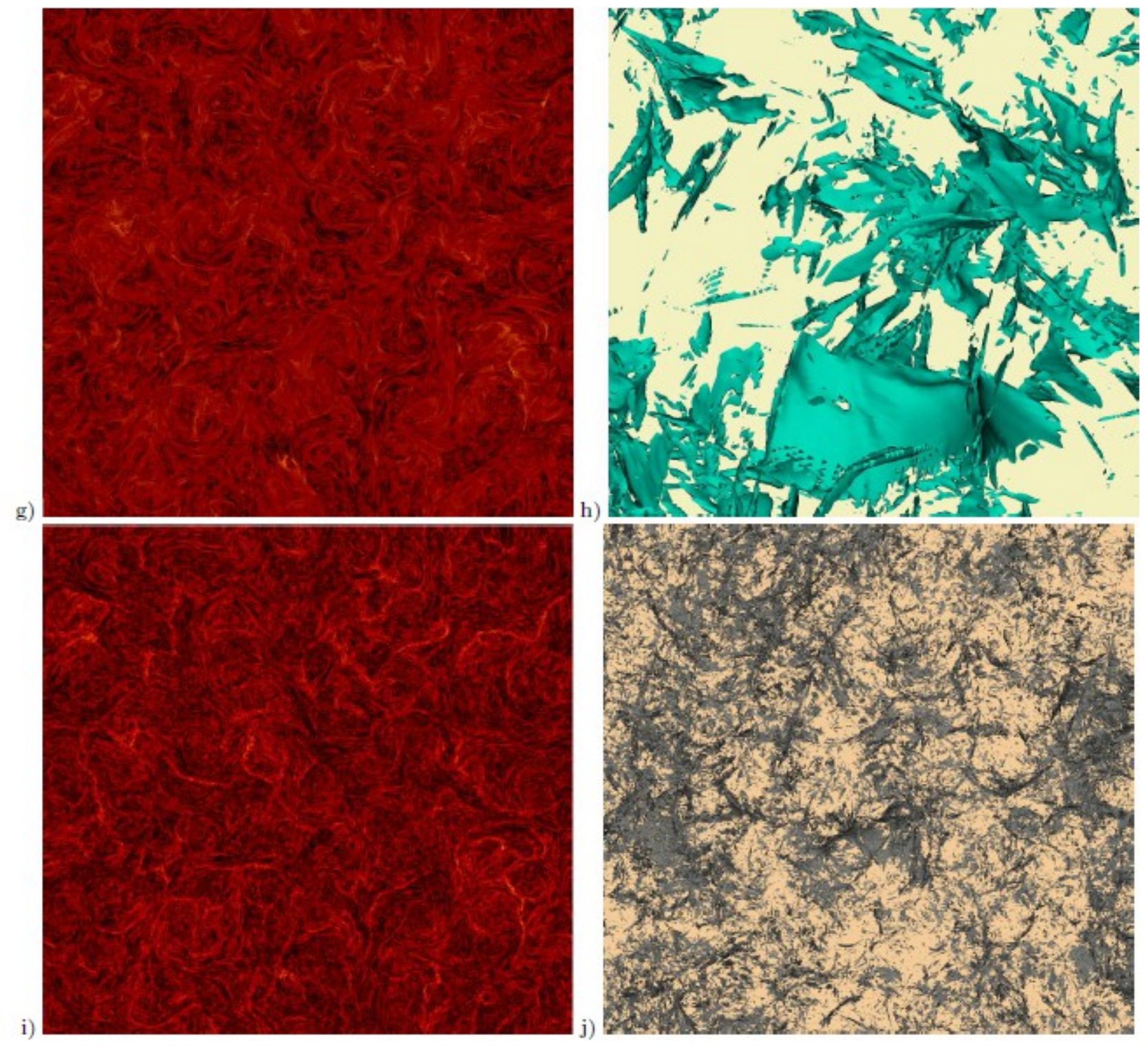}
\caption{Real space structures and iso-surfaces for various quantities in the decaying case at $t=9.2$. a) Cut across the plane of  magnetic field structure, b) iso-surfaces of the magnetic field (zoomed by 1.4 times), c) 3D-view of magnetic helicity structures, d) iso-surfaces of magnetic helicity, e) cut across the plane of current density, f) iso-surfaces of current density (zoomed by 12 times), g) cut across the plane of velocity field, h) iso-surfaces of velocity (zoomed by 14.4 times) i) cut across the plane of vorticity field and j) iso-surfaces of vorticity (zoomed by 12 times). Resolution in all these pictures is $1024^{3}$. (A color version of this figure is available in the online journal.)}
\label{fig3a}
\end{center}
\end{figure}

In the decaying case, the structures of various physical quantities are observed at time t = 9.7, when the peak of the magnetic helicity spectrum has moved to $\it{k}$ $\sim$ 3, from $\it{k}$ $\sim$ 70 . Figure (3a) shows the cut across the plane of the magnetic field where strong tangled field structures and some magnetic reconnection regions\footnote{Magnetic reconnection regions in this work imply the regions where two counter rotating magnetic vortices are seen separated by a dark region. In this dark region, perpendicular to the plane, current sheet is observed(not shown in these figures). Please see section 4.5.1 for some remarks and also Figure 7.} in the field are seen. Several intermittent scales (scales over which the field strength shows abrupt changes) are also observed. The iso-surfaces (Figure (3b)) appear like twisted flux tubes which are also the dissipative structures in the field. Magnetic helicity structures in three dimensions are shown in Figure (3c), some of them extending over the entire box and can be compared to kinetic helicity flux surfaces seen in 3D-HD (see \cite{che03}). The iso-surfaces shown in Figure (3d)., show possibly the largest structures of any physical quantity visualized in our simulations. Since magnetic helicity represents the linkage and twists in the magnetic field, it can be seen that the linkages in the field appear to form two distinct features e.g. huge eye-like structures (magnitude of magnetic helicity is shown in Figure (3c)) in some places and column like structures all around. Surprisingly, magnetic helicity structures in the forced case exhibit exactly the same features as are observed in the decaying case. This observation supports the idea that irrespective of whether a driving is present or not, the flux tubes get linked the same manner, in an isotropic, homogeneous, incompressible MHD flow. \\
 Figures (3e) and (3f) represent the structures of current density $j^2$. These structures have a large magnitude but their sizes are small and are thinly spread over the entire plane, which is a signature of intermittency. The iso-surfaces of $\bf{j}$ are thin sheet like. Once these sheets are formed, they do not appreciably change in size and remain almost constant throughout the entire evolution of the system playing a very significant role in the inverse cascade \citep{mulmal12} (see also the section on current density spectra below). The velocity and vorticity field structures (Figures (3g) and (3i)) are low in magnitude, thinly spread over the entire plane and also show intermittency. The vorticity structures have structural resemblance (not magnitudinal) to the structures of current density and appear to be similar to the ones seen in 2D-HD of \cite{Bof07}. The iso-surfaces of velocity (Figure (3h)) and vorticity (Figure (3j)) appear in sheet like configurations, with the iso-surfaces of vorticity and current density showing high spatial correlation similar to the one observed in a Taylor-Green MHD decaying turbulent flow (see \cite{pouq101}). Such a high level of spatial correlation, probably stems from the frozen-in property of the magnetic field, at high Reynolds number, which leads to the formation of current sheets where ever fluid portions with different magnetic field orientation are colliding. As the fluid is tightly coupled to the magnetic field this leads to the emergence of vortex sheets in close vicinity of the current sheets \citep{Bis00}.\\
 For the decaying case, in the limit of exact conservation of magnetic helicity, it is stated that magnetic helicity tends to be a constant w.r.t. time $\it{t}$ (the large eddy turnover time of the system ), instead of evolving over resistive times scales as in the driven case \citep{chr05}. It is also known that the magnetic energy has a power law of $t^{-0.5}$ at larger magnetic Reynolds numbers \citep{Bis99,chr05}, where $t < < t_R$. The inference that can be drawn from combining these two facts is that, it is plausible for the large-scale magnetic structures to emerge faster than $t_R$ in the decaying case. Our decaying system satisfies both the stated criterion (see Figure (9b) and its caption for more details).\\  
The magnetic field structures seen here are a result of build up of magnetic energy caused by the inverse spectral transfer of magnetic helicity as already stated in the above section. Although the initial values for kinetic energy and magnetic energy are the same, it is clearly seen that the magnetic field structures are stronger than their velocity field counterparts, at large scales. Similar comparison for the structures of magnetic helicity and vorticity (which is a good proxy for kinetic helicity as there appears to be a high degree of alignment between velocity and vorticity) shows that the magnetic helicity structures are the stronger counterpart.\\
\placefigure{fig3}
\placefigure{fig3a}
\subsection{Spectra of Current density}
The time lapsed current density spectra are useful in understanding the nature and evolution of the current filaments that form during the inverse cascade of magnetic helicity, with Figures (4a) (driven case) and (4b) (decaying case) representing this process. From these figures it is observed that the peak of current density spectra moves to relatively less larger scale(s) in both the cases, than the peak of the magnetic energy spectrum at approximately the same time (see for example: Figure (4a) and Figure (1a top)). This observation complements the visualized iso-surfaces of both magnetic field and current density, and is at the very heart of the inverse cascade phenomenon which could be realized as a merging of positively aligned and thus mutually attracting current carrying structures \citep{brem93, mulmal12}. In this process, merging of even small-scale current structures, can result in large-scale magnetic structures with an increase in magnitude of the current density (and not necessarily the formation of large-scale current filaments), which is observed in the spectrum as well in the structures. The transfer of energy from small scales to large scales via the interaction, linkage and twisting of the magnetic field lines of various scales can be attributed to this same process, thus making the inverse cascade of magnetic helicity highly non-local (see \cite{mulmal12}).

\begin{figure}
\begin{center}
a) \includegraphics[width=7cm,height=7cm,viewport=45 20 330 260,clip]{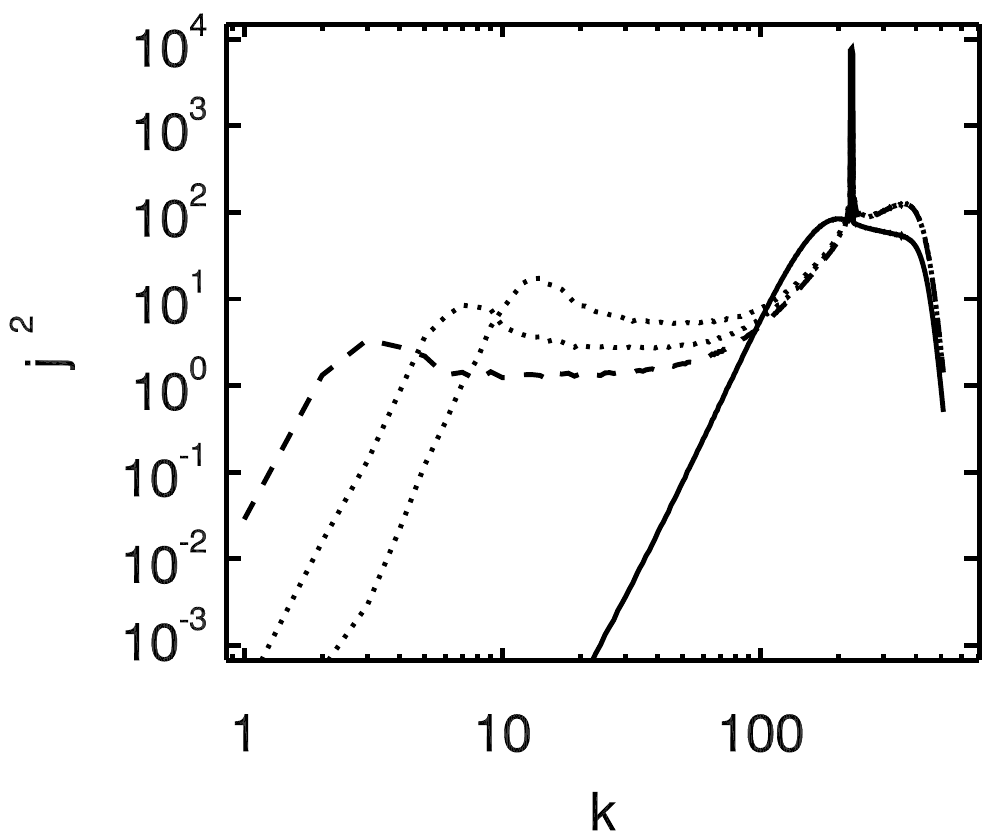}
b) \includegraphics[width=7cm,height=7cm,viewport=45 20 330 260,clip]{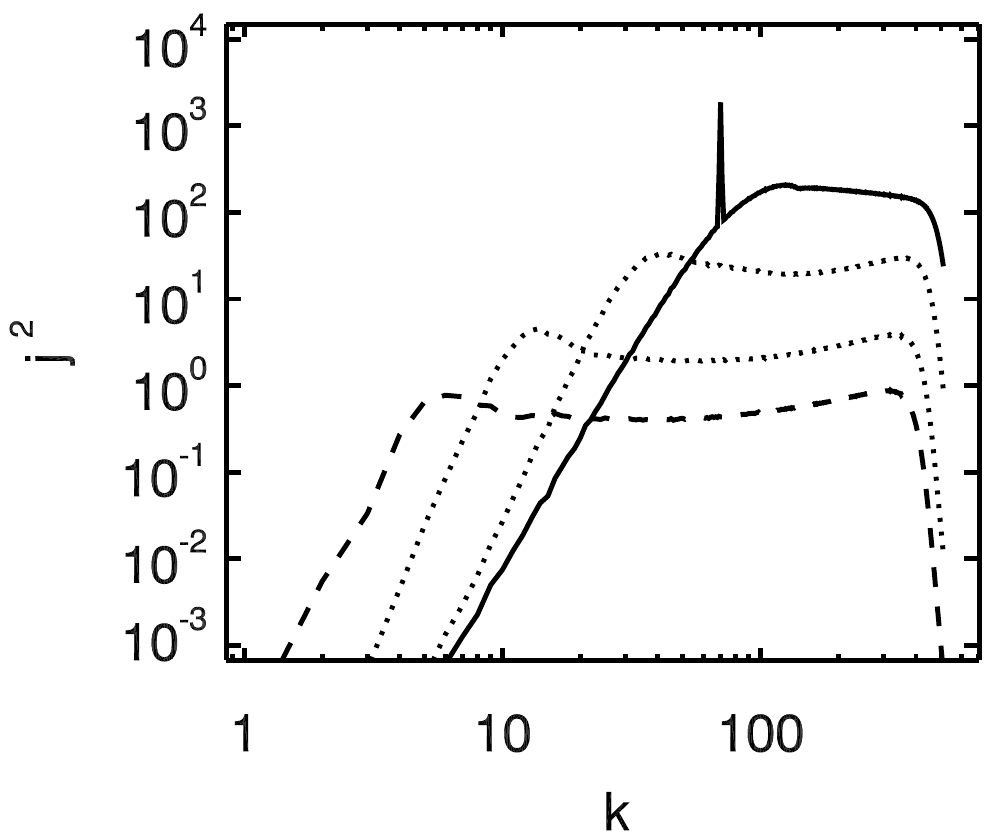}
\caption{Spectra of current density. a) Driven case shown at t = 0.17, 0.36, 5.05 and 6.66, b) decaying system shown at t = 0.30, 2.28, 6.78 and 9.30. Dark line: initial state, dotted lines: intermediate states, dashed lines: final state.}
\label{fig4}
\end{center}
\end{figure}

\placefigure{fig4}
\subsection{System initially forced and later allowed to decay}
\begin{figure}
\begin{center}
\includegraphics[width=0.9\textwidth,viewport=55 315 535 760,clip]{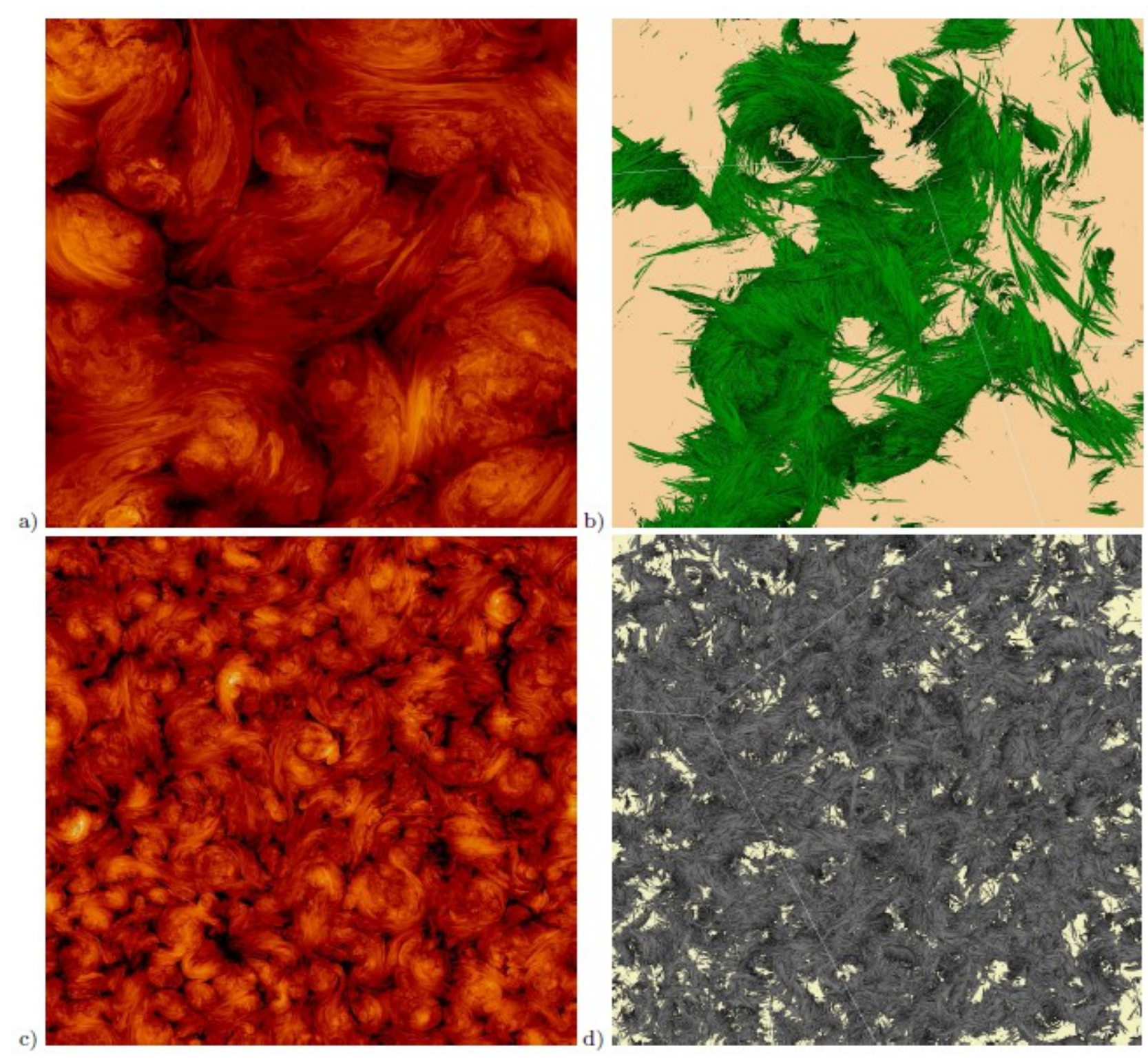}
\caption{System initially forced and later allowed to decay. a) Cut across the plane of magnetic field structures at t = 9.9 after the forcing was stopped at $t_3$ (see Figure (1a)) b) iso-surfaces of the magnetic field for Figure (a) (zoomed by 1.5 times), c) cut across the plane of magnetic field structures at t = 5.3, after the forcing was stopped at $t_2$ (see Figure (1a)) and d) iso-surfaces of the magnetic field for Figure (c) (zoomed by 1.7 times). (A color version of this figure is available in the online journal.)}
\label{fig5}
\end{center}
\end{figure}

Large-scale magnetic structures in astrophysical systems will neither be completely represented by forced MHD turbulence nor by decaying MHD turbulence alone. It can be expected that the MHD turbulence in these systems could be a consequence of both forced as well as decaying turbulences \citep{sow02}. Present day observations do not capture the forcing scales, though they do capture the forcing events. The intermittent scales and events have also not been observed in detail. Radio observations, do observe the final resultant large-scale magnetic structures (see for e.g. \citep{falpa03}). It is plausible that a system (say an isolated plasma) initially experiences a forcing (in the form of a shock wave or a jet or any such high energy event) through which it becomes MHD-turbulent and once the forcing is switched off, the same system will then become a decaying turbulent flow and over time might result in the observed large-scale magnetic structures. Till date there are no simulations which substantiate or capture such a behavior. Here we report an attempt which tries to mimic this behavior. For this, we perform three distinct decaying simulations, by stopping the forcing at three specific points in time ($t_1$,$t_2$ and $t_3$ in Figure 1) in the evolution of magnetic energy spectrum in the forced case (see Figure (1a)) and allowing the system to evolve as a decaying turbulent flow from that instance of time. This strategy allows us to achieve two goals i.e. 1) to validate the assumption that both forced and decaying turbulence are necessary for large-scale magnetic structure formation and 2) to look for any dependency on the size of the structures formed and the amount of time the system is forced. When the forcing is withdrawn from the system and is allowed to decay, the system represents a decaying turbulence system in which initial energy values of kinetic and magnetic energies differ. This system is thus a variation of the decaying turbulence case of the above section where the system was initialized with equal magnetic and kinetic energies. \\
 Each of these simulations was performed for a period amounting to $\sim$ 5 time units, as it was felt sufficient to clearly observe the necessary trends. Figures (5a) and (5b) represent the final state of the decaying simulation that resulted after the forcing was stopped at $t_3$. Similarly Figures (5c) and (5d) represent the final state for the resulting decaying simulation, with forcing stopped at $t_1$. From these figures it is clear that the time the system is forced has a bearing on the size of the structures formed. However, it is plausible that the structures from Figure (5c) will evolve into the same size as Figure (5a) provided they have enough energy as theoretically inverse cascade or inverse spectral transfer can proceed to infinitely large-scales. The structures at $t_2$ are an intermediate between these two cases and are not shown here.\\
The important observations from these simulations are: a) the power law values in the spectrum of all the quantities changed from their initial forced state values to the power law values of the decaying case [the spectrum not shown here], (e.g. magnetic helicity which was satisfying a power law value of -3.3 in the initial forced state changed to -3.6 in the decaying run). b) The number of magnetic reconnection regions decreases from Figure (5c) to Figure (5a) indicating that while the system has predominantly intermediate scales, there are large number of magnetic reconnection regions, but they decrease as these intermediate scales merge or change to large-scale structures via inverse cascade / spectral transfer. Further, since the time at which the forcing is stopped serves as the initial condition for these decaying cases, the resultant structures of Figure 5, emphasize the role of initial conditions on the final out come of the simulations.\\
To the best of our knowledge, this is the first time that in MHD turbulence simulations, a change of power law from forced case to decaying case is observed. This also coincided with the improvement in the dimensionality of the structures from fractal (1.5) (see Figure (2d)) to proper two dimensions (see Figures (5b) and (5d)), from the initial forcing state to the final decaying case. These observations reiterate the assertion that the kind of small-scale forcing used in our simulations inhibits the formation of coherent large-scale structures but decaying turbulence (initiated at any scale) helps in the formation of coherent two dimensional structures. Although results from magnetic field structures only are being reported here, such a change in dimensionality of the structures from the forced case to decaying case are seen in several other quantities too (e.g. velocity field).\\
\placefigure{fig5}
\subsection{Evolution of the Magnetic Field}
\begin{figure}
\begin{center}
\includegraphics[width=0.9\textwidth,viewport=60 315 530 750,clip]{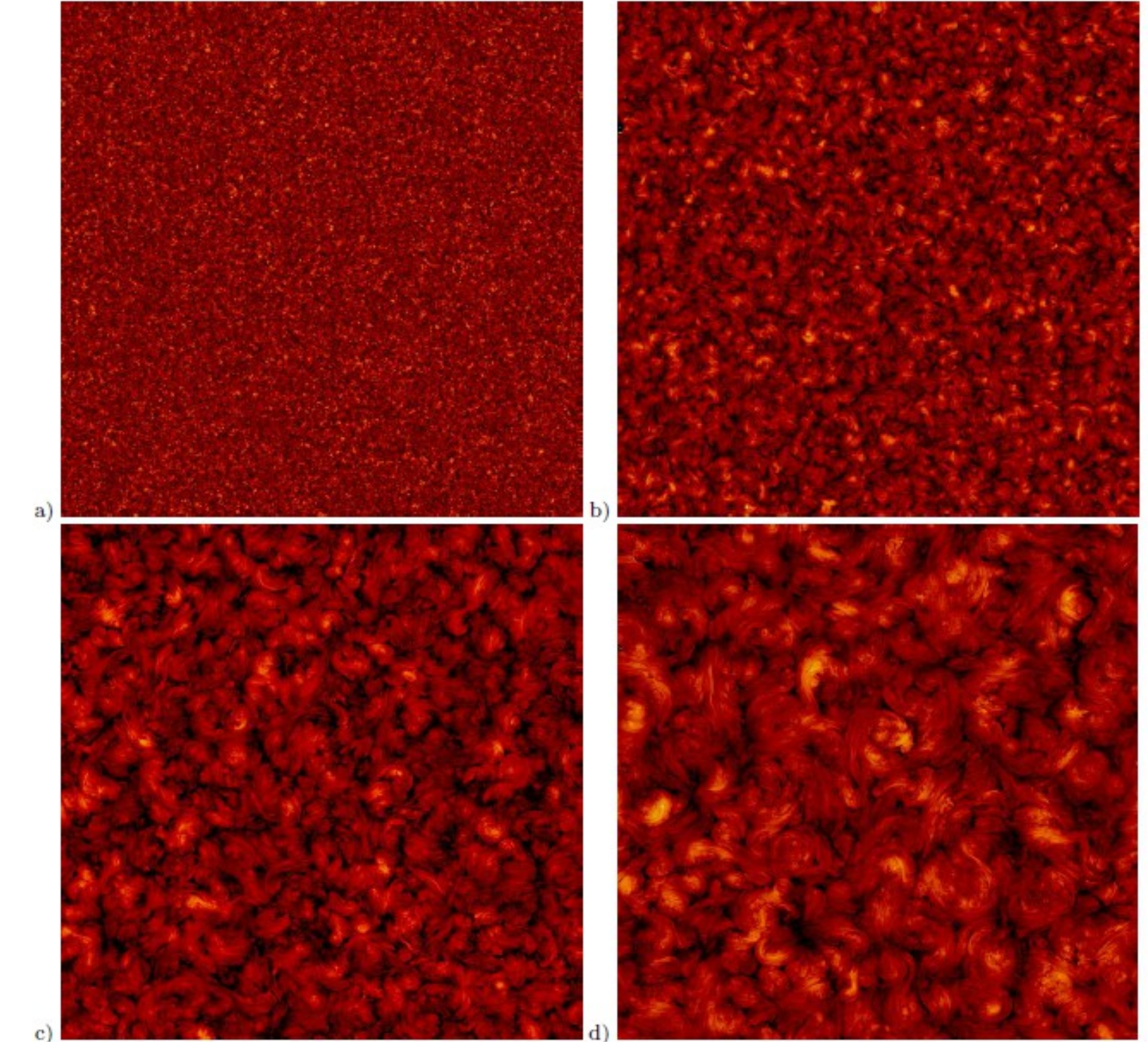}
\caption{Evolution of the magnetic field. a) Cut across the plane of  unfiltered magnetic field structure (at $t= 0.33$, where the forcing was stopped), b-d) cut across the plane of magnetic field structures (at $t=4.5,5.5$ and $8$). (A color version of this figure is available in the online journal.)}
\label{fig6}
\end{center}
\end{figure}

 The Figures (6a)-(6d) show the evolution of magnetic field from an initial random state to a state with coherent large scale structures for the decaying case.\\
 As decaying turbulence takes over, from the randomness of the initial state (Figure (6a)), structure formation is seen (Figures (6b)- (6d)), where in the initial, point like, random magnetic field structures grow into small vortex like structures. Here already several magnetic reconnection regions emerge. The process continues further and the structures grow larger as the reconnection regions become fewer in number. While this happens through the inverse spectral transfer in the decaying case, the system is also loosing energy. Thus the magnitude of the field strength in these structures continues to decrease, which can also be seen in Figure 2 of \cite{mul12}, where the peak of magnetic energy falls as $k^{-1}$. This can be seen from the increase of dark regions in these figures (note that in these pictures bright regions indicate higher field strength and dark regions are of lower field strength).\\
\placefigure{fig6}
\subsubsection{Magnetic reconnection and Magnetic helicity}
\begin{figure}\nonumber
\begin{center}
 \includegraphics[width=0.9\textwidth]{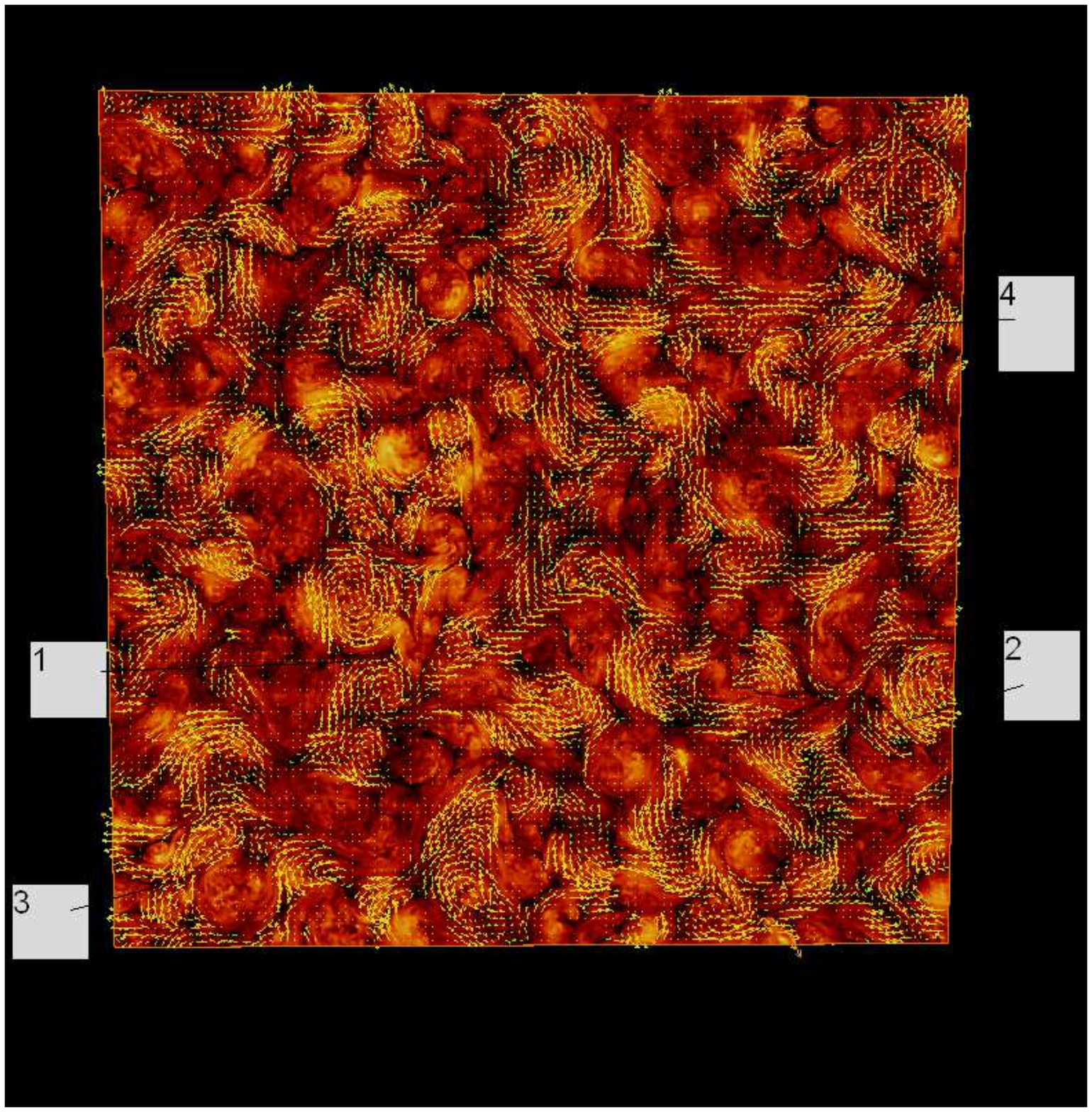}
\caption{Magnetic reconnection in the decaying case. Shown in the figure is a superimposition of magnetic field vectors on absolute magnetic field structures. Also shown are four reconnection regions marked 1 - 4. Resolution for these figures is $512^3$ with $\hat{\mu}_8$ $\sim$ $\hat{\eta}_8$ = 2e-35 and the shown figure is taken at $t=4$ with forcing stopped at $t=1$.  (A color version of this figure is available in the online journal.) }
\label{fig7}
\end{center}
\end{figure}

Generally, large-scale magnetic structures can only develop from small-scale structures through a change in field topology necessitating magnetic reconnection. This process requires that the uniqueness of 
magnetic field lines is lost on sufficiently small spatial scales. This can be accomplished, for 
example, by resistive diffusion or by spontaneous stochasticity present in rough turbulent fields.
The former reconnection paradigm (cf., e.g., the Sweet-Parker model) implies reconnection dynamics on 
resistive time scales, see, e.g.,  \citep{bis03}, and breaks magnetic-helicity conservation while 
the latter concept underlies the turbulent reconnection model yielding higher reconnection rates 
independent of the magnetic diffusivity, e.g. \citep{ey11} and references therein. Therefore, theoretical constraints relating the reconnection rate and the rate at which inverse cascade proceeds should be expected. 
However, the role of magnetic helicity in reconnection is not clearly understood, let alone the relationship between inverse cascade and reconnection, although there exists several studies (both 2D and 3D) in this regard \cite{ hey84,sch88,laz99,lap08,ser09,ser10,laz11,pri12}. Some of these recent studies put forward the concept of turbulent reconnection to comply with the astrophysical observations that show faster reconnection than earlier envisaged models (i.e. Sweet-Parker)(c.f. \cite{kow09,laz11}). If the reconnection rate indeed is faster, it shall have interesting consequences on the inverse cascade dynamics. However, an investigation of the influence of different reconnection physics is beyond the scope of this paper. Here in this work, we point out that while large-scale magnetic fields are generated through inverse cascade of magnetic helicity, it seems plausible that such a process proceeds through successive steps of turbulent reconnection events, although qualitative or quantitative analysis of such a mechanism, is currently beyond the scope of this work. These remarks are made in the light of the unique (simplistic) 3D-MHD flow and numerical setup used to obtain the magnetic field structures observed in Figures 3,5,6 and 7 that show various reconnection regions.\\
\placefigure{fig7}
\subsection{PDFs}
Scale-dependent probability density functions (PDFs) are a useful tool to investigate the formation
of magnetic structures evolving through the inverse cascade of magnetic helicity, as they are one of the important tools used in understanding features like intermittency and  coherence of structures in turbulent flows (see e.g. \cite{sor01,far03,hol06,sal09,yos12}). We use the PDFs $P(\delta_{l}b)$ of the increments of the magnetic field $\delta_{l} b$ over a 
distance $l=|\vec{l}|$, where
{$\delta_{l}b=\frac{1}{|\bf{l}|}\bf{l}\cdot
$(\bf{b}(\bf{x}+\bf{l}$,t)-$\bf{b}(\bf{x}$,t))$},
to study the magnetic field at different length scales. 
PDFs of $\delta_l b$ are shown in Figures (8a) to (8c)
for several $l$. 
A total of 32 spatial scales are considered. These cover distances ranging from $l=\Delta x$ to $100 \Delta x$ , where $\Delta x=\frac{2 \pi}{1024}$ is the scale of the numerical grid.
Thus these PDFs characterize fluctuations whose sizes range from almost a single grid point to 
$ \sim 1/10^{th}$ the linear extension of the periodic box. 
In all figures, they have been normalized to unit variance and shifted in amplitude for clarity.
In Figure 8, two different types of PDFs are observed: (i) Gaussian-like (for structures of Figures (2a) and (2b)) and (ii) PDFs with sharper peaks and broader tails showing clear deviation from Gaussianity (for structures of Figures (3a), (5a) and (5c)).\\
	The first variety of PDFs is represented by Figures (8a) and (8b). Figure (8a) represents a state very close to the initial states of both decaying and forced cases ( at $t_1$ of Figure (1a)). Figure 8b represents the final state of the forced case (close to $t_3$ of fig.1a). These profiles (Figure (8a) and (8b)) display Gaussian character on all spatial scales. This is in 
disagreement with known simulation results \cite{Mu00,ha04} of homogeneous MHD turbulence which exhibit a qualitative change of PDFs from large-scale Gaussian to small-scale leptocurtic shape as the underlying increments approach dissipative length scales. The observation indicates that the small-scale Gaussian forcing apparently dominates 
the two-point statistics of the system at smaller increment scales. This supports the argument that, in the driven case, emergence of larger scale spatial coherence, i.e. the build-up of long-range correlation, is inhibited by the forcing, as discussed in section 4.1.\\ 
The second type of PDFs, shown in Figure (8c), represents the final state of the decaying case or case 3
(initially driven and later allowed to decay). These PDFs are characterized by a sharp central peak and broader tails (wings) obtained from small to intermediate spatial distances becoming almost Gaussian at largest increments. 
These characteristics signal the appearance of some kind of `order' in the structures 
as opposed to the initial random state (compare Figures  (6a) to (6d)) in the decaying case ( and case 3). 
The PDFs indicate the presence of strong intermittent events. A plausible interpretation
of such a signature is the existence of various spatially coherent structures 
(cf., e.g., \cite{sal09}) which lead to intermittency signatures whenever a zone of coherence
is left or entered. At the small to intermediate bin measures and in agreement with the physical picture of merging of 
current filaments underlying the inverse cascade, the structures appear like twisted  flux tubes of similar sizes (see Figures (5b) and (5d)) resulting in strong correlations and enhanced coherence, a trend clearly captured in the profiles of Figure (8c). At the largest values of $l$, the PDFs tend to show Gaussian behavior as the involved two field values become increasingly de-correlated.\\ 
 It is interesting to note that the shape of the PDFs change from Figures (8a)-(8c) from the initial state of the decaying system to its final state. A similar trend is observed in case 3  where the system is initially driven and later allowed to decay.  If the system is allowed to decay longer, the broadening of the wings of the profiles of small to intermediate bin sized measures become more pronounced  (these curves are not shown here), (e.g for structures of Figure (5a)) indicating enhanced coherence among the structures. This observed trend could be verified  from such an astrophysical phenomenon where statistical data is available at both initially driven state and later at the decaying state (e.g. ARs of the Sun).    This probably, is one of the most important observation emerging from this studies.

\section{Discussion}
 The strongly idealized simulation setup chosen in this paper has some drawbacks e.g. finite size effects, slow resistive magnetic reconnection dynamics and limited scale-separation. This certainly reduces the realism of the simulations but not their potential astrophysical relevance as the turbulent inverse cascade of magnetic helicity studied in this paper (specifically in the decaying case and case 3) is a robust magnetic structure formation process. Thus it seems highly probable that the inverse cascade underlies the magnetic dynamics of physically more complex turbulent astrophysical systems.
 Hence, in following discussion, we interpret some of the observed astrophysical scenarios that closely resemble our simulation results from the case 3 (i.e. initially driven and later decaying) and the decaying case. It is important to note that inferences reached here, can only be approximate and not absolute, as modifications to the cascade process due to additional physics influencing the turbulent flow e.g. Hall or kinetic effects, can be expected and will be the subjects of future investigations.\\
 We also state here that our driven turbulence simulations need lot more modifications (as stated in section 4.1) to mimic the astronomical observations. 
\subsubsection{Identifying parameters to understand force-free structures}
Force free magnetic configuration \citep{tay74} is a concept 
that has received a lot of attention in both nuclear fusion studies 
and astrophysics \citep{bis03}. 
A rough estimate of the limiting size to which the magnetic structures evolve in the 
decaying turbulence setup described above can be obtained with the help of the 
dissipation rate of magnetic energy and the correlation length of the magnetic field.
which is a direct measure of the size of the structures. These quantities
characterize the decay of nonlinear turbulent dynamics and the size of large-scale magnetic 
coherent structures, respectively.\\
The magnetic correlation function is defined as:
\begin{equation}
\rho(\it{r})=\frac{\int^{X}_{-X}{\it{c_{x}}(\it{x})\cdot\it{c_{x}}(\it{x+r})\it{dx}}}{\int^{X}_{-X}{\it{c_{x}}^{2}(\it{x})\it{dx}}}
\end{equation}
where `X' is taken to be much greater than any characteristic length scale associated with the fluctuations in $\it{c_{x}}$. Here it represents the boundary of the simulation box. The parameter $\it{c_{x}}$, as mentioned above, is a general representation of any physical quantity which, in the context of this work, is the magnetic field $\bf{b}$. The correlation length is defined as the point where the function $\rho(\it{r})$ falls to $1$/$\it{e}$ of its peak value. 
The evolution of the magnetic field correlation is measured at three points in time during turbulence decay. 
It increases with time and indicates a rising degree of large-scale magnetic field coherence. 
Similarly, magnetic energy dissipation ($\epsilon_{M}= - \int_{\it{V}}{\it{dV}{\eta_n} {\bf{j}}\cdot\Delta^{n/2-1}{\bf{j}}}$) values for these three points were also obtained from the available spectral data.
 The data is as shown in the Table 1.
 \placetable{tab1}
This data approximated by a linear model (solid line in Figure (9a)) indicates a growing 
correlation length as turbulence decays. The curve intercepts the x-axis at a correlation length of 0.22 (that is $\sim$ 22 times the initial correlation length). This is roughly the maximal size of 
magnetic structure achievable in this simulation of decaying MHD turbulence.
The formed stable magnetic structure is regarded as approximately `force-free'. \\
 This approach is supported by observing the asymptotic behavior of the magnetic energy and its dissipation in the decaying turbulence case. Figure (9b) shows such a log-log plot,  of magnetic energy (solid curve) and its dissipation (dotted curve) versus the time, with power-law fits. The fits show that in the decaying MHD turbulent flow magnetic energy decays as $t^{-0.5}$, its dissipation rate follows dimensionally correct $t^{-1.5}$ curve for the most part of the simulation time.  It was observed by \cite{Mal09} that both kinetic energy dissipation and total energy dissipation ($\epsilon{_V}+\epsilon{_M}$) also follow the same power law of $t^{-1.5}$ . Thus the system stops dissipating both kinetically as well as magnetically quite rapidly and achieves after some additional transient relaxation; the anticipated force free magnetic state.\\
  Currently, our simulation is one-off in this direction, several such attempts are warranted in order to establish a statistical estimate of the ratio of the size of magnetic structures from their initial state to the final relaxed state.  
 \cite{kah13} show in their simulations that magnetic energy decays as $t^{-{\frac{2}{3}}}$ while the correlation length goes as $t^{{\frac{2}{3}}}$, in the evolution of primordial magnetic fields, through decaying helical MHD turbulence. Although, in our simulations, we do not observe these trends, the modeling described above could be helpful in improving our understanding of stable celestial (primordial) large-scale magnetic structures. 
\placefigure{fig9}
\subsubsection{Planetary magnetospheres}
 Several varieties of dynamo processes have been invoked to explain the generation of large-scale magnetic fields in planets (see \cite{caj11} and the references thereof). However, these dynamo mechanisms do not explain the existence and stability of large magnetospheres these planets possess. A plausible explanation is the inverse cascade of magnetic helicity which can transport magnetic energy into large scales from extremely small scales, as seen in these simulations and thus can result in the formation of stable coherent large-scale structures. Magnetic reconnection seen in the planetary magnetospheres, could be a natural ingredient in this large-scale structure formation.
 
 \subsubsection{Random small-scale component and coherent large-scale component}
 In astrophysical systems, observations detect the presence of two magnetic field components: random small scale magnetic structures (e.g. in galaxies \citep{bec96} or magnetic carpet in the Sun \citep{sti02}) and coherent large-scale component (in the same systems). In our simulations, the initial state of the both the decaying and the forced cases is similar to random small scale component. By the end of the driven simulation, a small amount of coherence is observed in the small-scale structures with the field still keeping a large amount of its randomness (an inference drawn from PDFs of Figures (8a) and (8b)). No clear coherent large-scale structures are seen (cf. sections 4.1 and 4.6). At the end of the decaying case simulation, the small-scale structures are both highly correlated and coherent as well, while at large scales appear to become uncorrelated . It is important to note that the emerging large-scale structures in decaying case are coherent, as observed from Figure (8c) as well as discussion in sections 4.2, 4.4 and 4.6. It is thus difficult to observe both the random small scales and coherent large scales, simultaneously, in these simulations. In astrophysical environments, these two scale ranges can co-exist because of the huge ratio of scale separation as well as time spans involved in the evolution of the large scales. With the available computing capacity and computing time, the tasks of (a) having several orders of scale separation or (b) to observe the emergence of coherent large scales by overcoming the random effects of small scales, appears to be a difficult ask (cf. section 4.1). It is important to insist here that one of the main purposes of the present simulations is to establish the role of decaying turbulence in the formation of coherent large scale magnetic structures in an initially driven system (or a system with significant amount of total energy and helicity in the initial state), which has hitherto not been studied in such great detail. 
\subsubsection{Large-scale magnetic structures in the  radio relics of galaxy clusters}
Origin of large-scale magnetic fields seen in the radio relics of clusters of galaxies is one of the open problems in astrophysics. Several works have tried to address this issue (see e.g. \cite{ens05, ens06, rus07}). Many of these attempts explain the way these large scale structures are seeded. Some of these seeding mechanisms include seeding through the galactic outflows during the starburst phase of galactic evolution \citep{don08}, galactic winds \citep{ber06} to name a few. From the observational data (see \cite{van09, bag09}) it is clear that the radio lobes or radio relics are at a farther distance from the central engine, usually merged galaxies, that (in many cases) is no longer active. These radio lobes which some times extend over `kpc' scales,  are nothing but non-thermal synchrotron emission and synchrotron emission is a direct indicator of presence of magnetic fields.\\
The numerical simulations so far \citep{ens05,ens06,rus07,don08} have used realistic data and initial conditions to simulate the formation of such large-scale magnetic structures, but the problem remains largely open. Recently using analytical and numerical methods \cite{jwai10} have shown that magnetic relaxation and magnetic reconnection play a major role in the inflation process of these bubbles. Their work explains how a radio bubble moves and expands in the intracluster medium without breaking up. However, it does not explain how large-scale structures evolve in the radio bubbles. The present work, although the simulation model is more idealized by the assumption of spatial homogeneity of turbulence, attempts at explaining the formation of large-scale structures.\\
The forcing used in our simulations produces tangled (helical) magnetic fields, which are seeding small scale magnetic fluctuations and significant small scale turbulent fluctuations in the velocity field. Such tangled magnetic fields have been observed in the jets that originate from the merger of galaxy clusters \citep{gab12}. Evidence of MHD turbulence in the intra-cluster plasma is present both in observations \citep{schu4,cl06a,cl06b,van09,cas11} as well as in simulations \citep{ens02,ens09b}. At these seed scales, there is a transfer of kinetic energy to magnetic energy, which then is transported via the inverse cascade of magnetic helicity to larger scales as observed in the present simulations. Due to the large amount of energy present in the helical jet, it is plausible that during its travel, the plasma at farther distances from the central engine, which was hitherto non-turbulent, becomes turbulent (i.e. seeding turbulence at new scales). When the forcing stops (much like the jets stop emanating from the central merger portion after a certain period of time), there is significant energy at these newly seeded scales which starts decaying. Since this decay process is taking place in a turbulent plasma, the concepts of decaying MHD turbulence i.e. inverse spectral transfer of magnetic helicity flux and formation of large scale structures come into play and can produce large-scale homogeneous magnetic fields observed in the radio bubbles. Previously the formation of such large-scale magnetic structures via the inverse cascade mechanism has been described as speculative since no compelling mechanism is known for generating large net helicity in a cosmic fluid \citep{wi02}. Recently, theories on (a) the production of magnetic helicity in hot lepton plasma in the early Universe \citep{semi07} and (b) production of primordial magnetic helicity through electro-weak phase transitions (see \cite{chr05,sub10,kah13} and references therein), have been proposed. In the wake of these new ideas [especially the ones mentioned in (b)], inverse cascade is a plausible mechanism to generate large-scale magnetic fields at cosmic scales \citep{chr05,sub10}. The roles of other important quantities such as kinetic helicity and the ratio of energies (at large scales, this ratio is $\sim$ 10) in the inverse cascade process have been made evident in the direct numerical simulations presented in \cite{mul12, mulmal12}. These results, in combination with the discussion on evolution of large-scale coherent magnetic structures from initial small scales presented above (see section 4.5), make a strong case in favor of inverse cascade of magnetic helicity as a mechanism for large-scale magnetic structure generation. In short, the jets (with probable helical magnetic fields) emerging out of a cluster merger, are inflating these radio `bubbles', which then rise and expand towards the outskirts of the cluster. But, the expansion of the bubble by itself might not be sufficient to explain the very coherent magnetic field observed in these bubbles and invoking the above explained inverse cascade mechanism is one plausible solution, as the jet magnetic field will have seed helicity and this might drive the process towards ordering the field within the bubble, in a turbulent environment \citep{ens09}, once the jet gets switched off. Since our simulations also point to the possibility that as the size of the structures increase, magnetic dissipation decreases and that the structures tend to move towards a force free stable configuration; these simulations also can in principle offer a natural explanation to the stability of these huge magnetic structures.
The dominance of small scales in the forced case observed in the simulations are not seen in the current day astronomical observations  because of the lack of sensitivity of the instruments to such small scale magnetic fluctuations. 
\subsubsection{Understanding small to intermediate scaled magnetic structures}
When the sensitivity of the astrophysical instruments detecting magnetic fields increases, they might be able to detect intermediate to small sized magnetic structures. It is plausible that formation and evolution of such scales could be interpreted through our simulations from case 3, where it is shown that the size of the structures formed depended on the point of stoppage of the force, before  the system starts decaying. \\
 Stochastically modeled observational data of either size of the structures or the number of reconnection regions or both might give insights on the nature and influence of the source that generated these structures, as there appears to be a direct relation between the number of reconnection regions and size of the structures formed through inverse cascade (see Section 4.4). A quantitative  form of such a relation, though, is beyond the scope of this  work. In addition, just one numerical forcing method as in the present paper is not sufficient for stochastic modelling and requires several possible data sets obtained from different forcing methods. The work presented here, thus, offers a good starting point for further such studies.
\subsubsection{A word of caution}
  It is possible that in the distant past, a strong force (e.g. energy imparted by a super nova explosion) acted on an isolated plasma for a very small amount of time. The plasma, then could be in decaying turbulence mode for a very long time that might result in extremely large-scale structures. Thus, while interpreting the properties of the observed large-scale structures using the concepts from case 3  presented above, one cannot ignore this importat possibility. If ignored, it  can lead to incorrect conclusions about the nature and influence of the source itself, for some observations. However, to simulate this scenario using direct numerical simulations, spatial resolutions many times larger than the ones used here and much smaller time steps are necessary. Owing to the limited computing resources, we could not perform this simulation but cannot rule out cautioning about this plausible scenario here, for completeness.\\
 The discussion in this section might also be helpful in understanding magnetic structures in the interstellar and intergalactic medium, which are also highly turbulent.\\
 \section{Conclusions}
A set of homogeneous forced and decaying 3D-MHD turbulence simulations were conducted. The forcing mechanism and initial conditions were placed in the small scales in the forced case and intermediate scales in the decaying case. The primary property that was studied closely was the inverse cascade of magnetic helicity and the large-scale magnetic structure formation due to the same. The important conclusions from the discussions in the above sections are:\\ 
1) It is observed that in the forced case coherent large-scale magnetic structures are not formed where as in the decaying case such structures are indeed generated. If the forcing is stopped and the system is allowed to decay, from the incoherent large-scale structures of forced case, coherent large scales evolve.\\
2) If the simulations are performed with large enough scale separation, stage by stage evolution of the magnetic structures from small scales to large scales via intermeidate scales can be observed, as shown in our work above. Hence it is suggested that the forcing mechanisms and initial conditions used in several other studies (see e.g. \cite{Bra01, min06}), if appropriately modified to achieve the necessary scale separation, would help in a better understanding of the process of evolution of large-scale structures from small scales. Such simulations will further help in minimizing the unphysical dependence on different initial conditions and forcing mechanisms.\\
3) The approach to form large-scale magnetic structures using both the forced and decaying 3D-MHD turbulences resemble closely the astrophysical scenario such as large-scale magnetic structures seen in the radio bubbles or relics of the cluster of galaxies. In those systems, it is possible that the forcing mechanism that is responsible for the formation of the large-scale magnetic structures must have acted for a very short period of time in comparison to the life time of these structures it self. The forcing mechanism might impart energy into the plasma and it is also possible that the methods by which this energy is imparted would also stir up the magnetic helicity. Such tangled magnetic fields can decay, as shown in the simulations above, to form large-scale magnetic structures due to the inverse cascade of magnetic helicity.\\
4) These simulations present a plausible idealized theoretical framework for the formation of large-scale magnetic structures. Using mean magnetic field as an initial condition would make the simulations anisotropic. Further effects like rotation, convection, thermal pressure and gravitation can be added to these simulations to make them more realistic. Adding these additional ingredients into the model will definitely violate  the conservation of  magnetic  helicity, however these are essential logical extensions to make the simulations more realistic. To match the results with the real data from observations, further modifications and fine tuning of parameters are necessary so that they tend to agree with the observations both quantitatively and qualitatively. Such efforts will form the basis of future work.\\

\hspace{2.5cm} {\bf{Acknowledgements}} \vspace{3mm}

SKM wants to thank Dr. A.Busse, Dr. P.N.Maya, Dr. S.Ramya and Prof. C.Sivaram for their constant support and help in the preparation of the manuscript. The authors want to thank Prof. T.A. En{\ss}lin, Dr. Klaus Dolag and Dr. Alexis Finoguenov for fruitful discussions on the large-scale magnetic structure formation processes in the clusters of galaxies. They also want to gratefully acknowledge Prof A.Pouquet and Prof. P.Mininni for useful discussions on the interpretation of reconnection and correlation length results presented in this paper.

\clearpage

\begin{figure}
\begin{center}
\includegraphics[width=0.9\textwidth,viewport=80 130 510 725,clip]{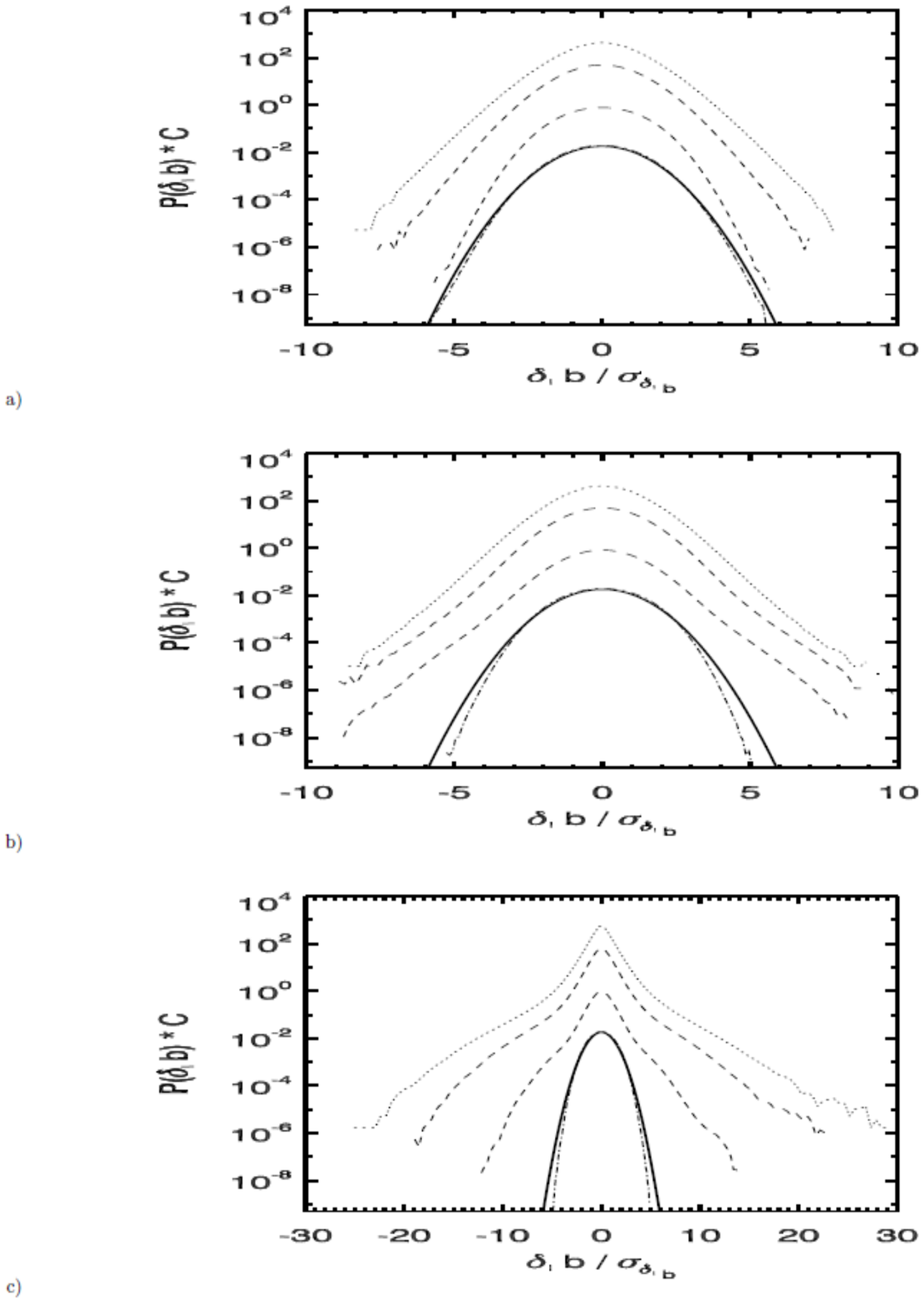}
\caption{PDFs of the magnetic field. a) Curves represent the initial state of the forced case / initial state of the decaying case b) curves representing the final state of the forced case and  c) Curves represent the final state of the decaying case / final state of the system(s) that is (are) initially driven and later allowed to decay. In all the three figures: dash-dot curves: plots obtained by measuring the field increment using the `largest bin size'. Bin size decrease from curves plotted with dash-dot curves to  curves plotted with `dotted line', which is the outer most curve. A model Gaussian with unit variance is also shown (solid black line curve).}
\label{fig8}
\end{center}
\end{figure}

\begin{figure}
\begin{center}
\includegraphics[width=0.9\textwidth,viewport=80 312 520 760,clip]{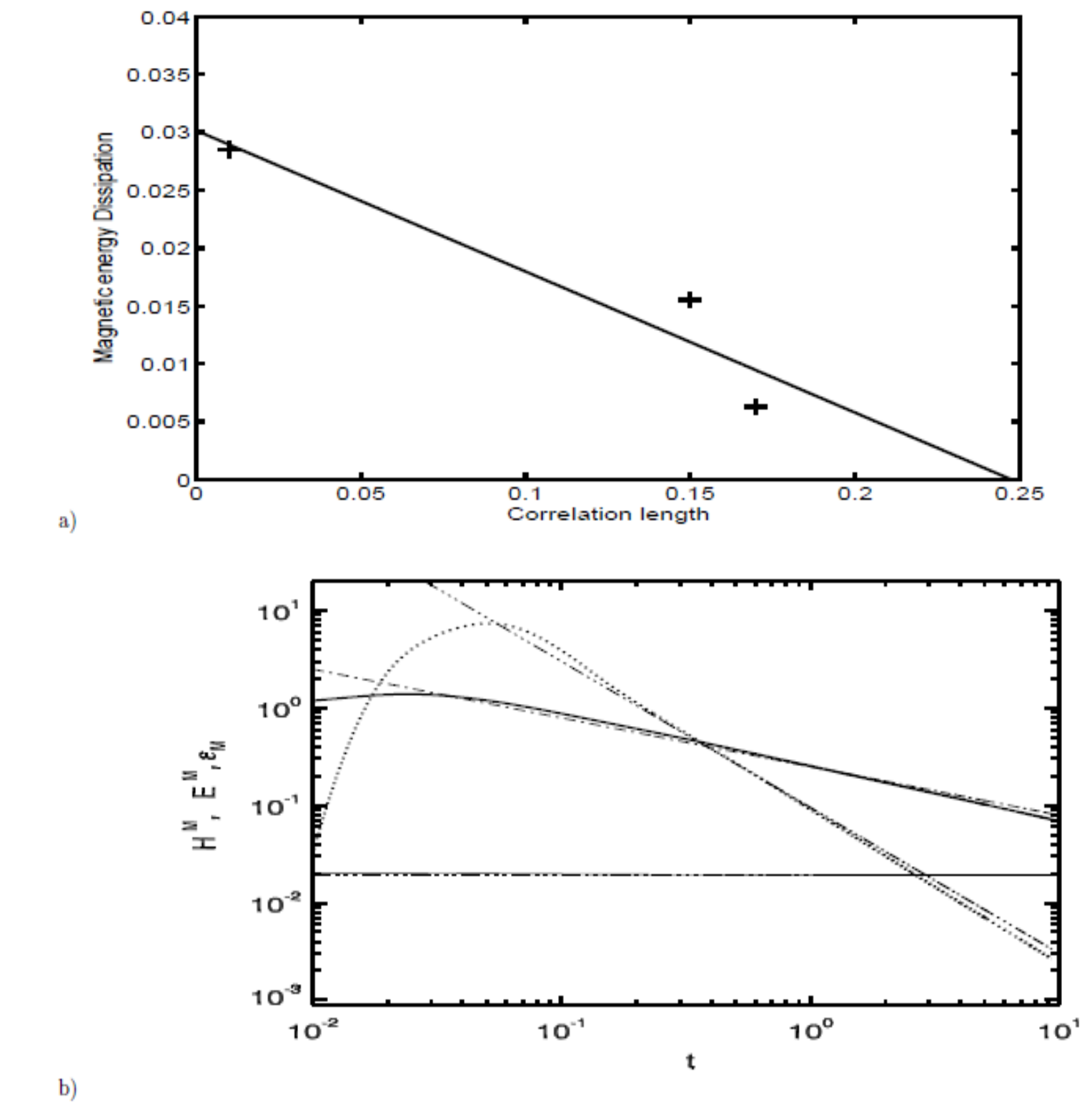}
\caption{Correlation length Vs magnetic energy dissipation and time evolution of magnetic energy and its dissipation and magnetic helicity in the decaying turbulence. a) Data points of Table. 1 (shown with a '+' sign) are fitted with a linear curve fit and b) magnetic energy and its dissipation in the decaying turbulence. Solid line: magnetic energy, dashed line: energy dissipation, dash-dot line: curve fit $t^{-0.5}$ to magnetic energy, dash-dot-dot line: curve fit $t^{-1.5}$ for the magnetic dissipation, thin continuous horizontal line (bottom of the figure) magnetic helicity and accompanying dash-dot horizontal line a curve fit with $0.02* t^0$ to emphasize that magnetic helicity is finite but constant in time, for this simulation.}
\label{fig9}
\end{center}
\end{figure}

\begin{table}\small
\begin{center}
\begin{tabular}{|c|c|c||}
\hline
Time& Length& Dissipation\\ \hline
t=0.233&0.01&0.0285\\ \hline
t=4.25&0.15&0.0155\\ \hline
t=5.89&0.17&0.0063\\ \hline
\end{tabular}\vspace{2mm}\\
\caption{Correlation lengths and energy dissipation values at different times for magnetic field structures} 
\label{tab1}
\end{center}\normalsize
\end{table}

\begin{thebibliography}{}
\bibitem[Alexakis et al.(2006)]{al06} Alexakis, A., Mininni, P., \& Pouquet, A., 2006, \apj, 640, 335
\bibitem[Alexakis et al.(2007)]{al07} Alexakis, A., Mininni, P., \& Pouquet, A., 2007, New Journal of Physics, 9, 298
\bibitem[Bagchi et al.(2009)]{bag09} Bagchi, J., Jacob, J., Gopal-Krishna, Werner, N., Wadnerkar, N., Belapure, J., Kumbharkhane, A. C., 2009, \mnras, 399, 601
\bibitem[Beck et al(1996)]{bec96} Beck, R., Brandenburg, A., Moss, D., \& Shukarov, A., 1996, Annu. Rev. Astron. Astrophys., 34, 155
\bibitem[Belenkaya(2009)]{Bel09} Belenkaya, E.S., 2009, PHYS-USP, 52, 8, 765
\bibitem[Bertone et al.(2006)]{ber06} Bertone, S., Vogt, C., \& En{\ss}lin, T.A., 2006, \mnras, 370, 319
\bibitem[Biskamp\& Bremer(1993)]{brem93} Biskamp, D., \& Bremer, U., 1993, Phys.Rev.Lett, 72, 3819
\bibitem[Biskamp \& M\"uller(1999)]{Bis99} Biskamp, D., \& M\"uller, W.-C., 1999, \prl, 83, 2195
\bibitem[Biskamp \& M\"uller(2000)]{Bis00} Biskamp, D., \& M\"uller, W.-C., 2000, Phys.of Plasmas, 7, 12, 4889
\bibitem[Biskamp(2003)]{bis03} Biskamp, D., 2003, Magnetohydrodynamic Turbulence (Cambridge: Cambridge University Press)
\bibitem[Boffetta(2007)]{Bof07} Boffetta, G., 2007, J. Fluid Mech., 589, 253
\bibitem[Brandenburg(2001)]{Bra01} Brandenburg, A., 2001, \apj, 550, 824
\bibitem[Brandenburg et al.(2002)]{Bra02} Brandenburg, A., Dobler, W., \& Subramanian, K., 2002, Astron.Nachr./AN, 323, 99
\bibitem[Brandenburg \& Sandin(2004)]{Bra04} Brandenburg, A., \& Sandin, C., 2004, A\&A, 427, 1, 13
\bibitem[Brandenburg \& Subramanian(2005)]{Bransub05} Brandenburg, A., \& Subramanian, K., 2005, Physics Reports, 417, 1
\bibitem[Braithwaite(2010)] {jwai10} Braithwaite, J., 2010, MNRAS, 406, 2, 705
\bibitem[Chen et al.(2003)]{che03} Chen, Q., Chen, S., \& Eyink, G.L., 2003, Phys. Fluids, 15, 2, 361
\bibitem[Canuto et al.(1988)]{can88} Canuto, C., Hussaini, M.Y., Quarteroni, A., Zang, Th.A., 1988, Spectral Methods in Fluid Mechanics (Berlin, New York: Springer)
\bibitem[Cassano et al.(2011)]{cas11} Cassano, R., Brunetti, G., \& Venturi, T., 2011, J.Astrophys.Astr., 32, 519
\bibitem[Christensson et al.(2005)]{chr05} Christensson, M., Hindmarsh, M., \& Brandenburg, A., 2005, Astron. Nachr. / AN, 326, 6, 393
\bibitem[Clarke \& En{\ss}lin(2006a)]{cl06a} Clarke, T.E, \& En{\ss}lin, T.A., 2006, \apj, 131, 2900
\bibitem[Clarke \& En{\ss}lin(2006b)]{cl06b} Clarke, T.E, \& En{\ss}lin, T.A., 2006, Astron. Nachr. / AN, No: 5/6, 553
\bibitem[Contopoulos et al.(2009)]{con09} Contopoulos, I., Christodoulou, D.M., Kazanas, D., \& Gabuzda, D.C., 2009, \apj, 702:L148 - L152
\bibitem[Cowling(1981)]{cow81} Cowling, T.G., 1981, Ann.Rev.Astron.Astrophys, 19:115-35
\bibitem[Donnert et al.(2008)]{don08} Donnert, J., Dolag, K., Lesch, H., \& M\"uller, E., 2009, \mnras, 392, 1008
\bibitem[En{\ss}lin \& Br\"uggen(2002)]{ens02} En{\ss}lin, T.A., \& Br\"uggen, M., 2002, \mnras, 331, 1011
\bibitem[En{\ss}lin et al.(2005)]{ens05} En{\ss}lin, T.A., Vogt, C., \& Pfrommer, C., 2005, The Magnetized Plasma in Galaxy Evolution, Proceedings of the conference held in Krakow, Poland, Sept. 27th - Oct. 1st, 2004, Edited by K. Chy\'zy, K. Otmianowska-Mazur, M. Soida, and R.-J. Dettmar, Jagiellonian University, Krakow 2005, p. 231-238
\bibitem[En{\ss}lin \& Vogt(2006)]{ens06} En{\ss}lin, T.A., \& Vogt, C., 2006, A\&A, 453, 447
\bibitem[En{\ss}lin \& Dolag(2009)]{ens09} En{\ss}lin, T.A., \& Dolag, K., 2009, Personal communication
\bibitem[En{\ss}lin et al.(2009)]{ens09b} En{\ss}lin, T.A., Clarke, T., Vogt, C., Waelkens, A., \& Schekochihin, A.A., 2009, RevMexAA (Serie de Conferencias), 36, 209
\bibitem[Eyink et al.(2011)]{ey11} Eyink, G.L., Lazarian, A., \& Vishniac, E.T., 2011, \apj, 743, 51
\bibitem[Falgarone \& Passot Eds.(2003)]{falpa03} Falgarone, E., \& Passot, T. (Eds.), 2003, Turbulence and magnetic fields in astrophysics (Berlin: Springer)
\bibitem[Farge et al(2003)]{far03} Farge, M., Schneider, K., Pellegrino, G., Wray, A.A., \& Rogallo, R.S., 2003, Physics of Fluids, 15, 10, 2886
\bibitem[Field \& Carroll(2000)]{fie00} Field, G.B., \& Carroll, S.M., 2000, Phys. Rev.D, 62, 103008
\bibitem[Gabuzda et al.(2012)]{gab12} Gabuzda, D.C., Christodoulou, D.M., Contopoulos, I., \& Kazanas, D., 2012, Journal of Physics: Conference Series, 355, 012019
\bibitem[Gilbert \& Sulem(1989)]{gil89} Gilbert, A.D., \& Sulem, P-L., 1989, Geophysical \& Astrophysical Fluid Dynamics, 51:1-4, 243-261
\bibitem[Graham et al.(2011)]{john11} Graham, P.J., Mininni, P., \& Pouquet, A., 2011, Phys. Rev. E, 84, 016314
\bibitem[Haugen \& Brandenburg(2004)]{ha04} Haugen, N. E. L., \& Brandenburg, A., 2004, Phys. Rev. E, 70, 036408 
\bibitem[Heyvaerts \& Priest(1984)]{hey84} Heyvaerts, J., \& Priest, E.R., 1984, A\&A, 137,63
\bibitem[Homann(2006)]{hol06} Homann, H., 2006, PhD thesis, Ruhr-Universit\"at Bochum
\bibitem[Jones(2011)]{caj11} Jones, C.A., 2011, Annual Review of Fluid Mechanics, 43, 583.
\bibitem[Kahniashvili et al.(2013)]{kah13} Kahniashvili, T., Tevzadze, A.G., Brandenburg, A., \& Neronov, A., Phys. Rev. D, 2013, 87, 083007
\bibitem[Kinney et al.(1995)]{kin95} Kinney, R., McWilliams, J.C., \& Tajima, T., 1995, Phys. Plasmas, 2, 3623
\bibitem[Kowal et al.(2009)]{kow09} Kowal, G., Lazarian, A., Vishniac, E. T., \& Otmianowska-Mazur, K., 2009, APJ, 700, 63
\bibitem[Lapenta(2008)]{lap08} Lapenta, G., 2008, Phys. Rev. Lett., 100, 235001
\bibitem[Lazarian \& Vishniac (1999)]{laz99} Lazarian, A., \&  Vishniac, E.T., 1999, APJ, 517, 700
\bibitem[Lazarian et al.(2012)] {laz11} Lazarian, A.,  Eyink, G.L., \& Vishniac, E.T., 2012, Phys. Plasmas, 19, 012105
\bibitem[Malapaka(2009)]{Mal09} Malapaka, S.K., 2009, PhD thesis, University of Bayreuth 
\bibitem[Malapaka \& M\"uller(2013)]{mal13} Malapaka, S.K.,  \& M\"uller, W.-C., 2013, APJ, 774, 84
\bibitem[Moffatt(1969)]{Mof69} Moffatt, H.K., 1969, J. Fluid Mech., 35, 117
\bibitem[Moffatt(1978)]{Mof78} Moffatt, H.K., 1978, Magnetic field generation in electrically conducting fluids (Cambridge: Cambridge University Press)
\bibitem[Miller et al(1996)]{mil96} Miller, R.S., Mashayek,  F., Adumitroaie, V., \& Givi, P., 1996, Phys. Plasmas, 3(9), 3304
\bibitem[Mininni et al.(2006)]{min06} Mininni, P., Pouquet, A., \& Montgomery, D.C., 2006, Phys. Rev. Lett., 97, 244503
\bibitem[Mininni(2007)]{min07} Mininni, P., 2007, Phys.Rev.E., 76, 026316
\bibitem[Mininni \& Pouquet(2009)]{min10} Mininni, P., \& Pouquet, A., 2009, Phys. Rev. E, 80, 025401(R)
\bibitem[M\"uller \& Biskamp(2000)]{Mu00} M\"uller, W.-C., \& Biskamp, D., 2000, Phys. Rev. Lett., 84, 475
\bibitem[M\"uller \& Grappin(2005)]{mulgra} M\"uller, W.-C., \& Grappin, R., 2005, Phys. Rev. Lett, 95, 114502.
\bibitem[M\"uller \& Malapaka(2008)]{Mu08} M\"uller, W.-C., \& Malapaka, S.K., 2008, Abstract: NG44A-012008, Fall Meet. Suppl., Eos Trans. AGU, 89(53)
\bibitem[M\"uller \& Malapaka(2010)]{mulmal10} M\"uller, W.-C., \& Malapaka, S.K., 2010, ASP Conference Series, N.V. Pogorelov, E. Audit and G.P. Zank (Eds.), 429, 28
\bibitem[M\"uller et al.(2012)] {mul12} M\"uller, W.-C., Malapaka, S.K., \& Busse, A., 2012, Phys. Rev. E, 85, 015302(R)
\bibitem[M\"uller \& Malapaka(2013)]{mulmal12} M\"uller, W.-C., \& Malapaka, S.K., 2013, Geophysical \& Astrophysical Fluid Dynamics, 107, 12, 93, DOI:10.1080/03091929.2012.688292
\bibitem[Passot et al.(1999)]{pas99} Passot, T., V\'azquez-Semadeni, E., \& Pouquet, A., 1999, \apj, 455, 536
\bibitem[Pouquet et al.(1976)]{Po76} Pouquet, A., Frisch, U., \& Leorat, J., 1976, J. Fluid Mech., 77, 321
\bibitem[Pouquet et al.(2010)]{pouq10} Pouquet, A., Brachet, M-E., Lee, E., Mininni, P., Rosenberg, D., \& Uritsky, V., 2010, Proceedings IAU Symposium $\bf {271}$, N.H.~Brummell, A.S.~Brun, M.S.~Miesch and Y.Ponty (Eds.),304.
\bibitem[Pouquet et al.(2010)]{pouq101} Pouquet, A.,  Lee, E., Brachet, M-E., Mininni, P., \& Rosenberg, D., 2010, Geophysical \& Astrophysical Fluid Dynamics, 104, 2-3, 115
\bibitem[Priest(2012)]{pri12} Priest, E.R., 2012, Magnetic Reconnection, Personal communication
\bibitem[Ruszkowski et al.(2007)]{rus07} Ruszkowski, M., En{\ss}lin, T.A., Br\"uggen, M., Heinz, H., \&  Pfrommer, C., 2007,  \mnras, 378, 662
\bibitem[Salem et al(2009)]{sal09} Salem, C., Mangeney, A., Bale, S. D., \& Veltri, P. 2009, \apj, 702, 537S
\bibitem[S\'anchez-Salcedo et al.(1999)]{san99} S\'anchez-Salcedo, F.J., Brandenburg, A. \& Shukurov, A., 1999, Astrophysics and Space Science, 263, 1-4, 87
\bibitem[Schekochihin \& Cowley(2007)]{sce07} Schekochihin, A.A., \& Cowley, S.C., 2007, Magnetohydrodynamics: Historical Evolution and Trends, 85
\bibitem[Schindler et al.(1988)]{sch88} Schindler, K., Hesse, M., \& Birn, J., 1988, J.Geophys.Res., 2, 247
\bibitem[Schuecker et al.(2004)]{schu4} Schuecker, P., Finoguenov, A., Miniati, F., B\"ohringer, H., \& Briel, U.G., 2004, A\&A, 426(2), 387
\bibitem[Semikoz \& Sokoloff(2007)]{semi07} Semikoz, V.B., \& Sokoloff, D.D., 2007, Tests of New Physics in Rare Processes and Cosmic Rays (Elementary Particles and Fields. Theory), Nuclear Physics (Russian), vol. 70, No. 1, 156
\bibitem[Servidio et al(2009)]{ser09} Servidio, S., Matthaeus, W.H., Shay, M.A., Cassak, P.A., \& Dmitruk, P., 2009, Phys. Rev. Lett, 102, 115003
\bibitem[Servidio et al(2010)]{ser10} Servidio, S., Matthaeus, W.H., Shay, M.A.,  Dmitruk, P., Cassak, P.A., \& Wan, M.,  2010, Phys. Plasmas, 17, 032315
\bibitem[Solanki et al.(2006)]{sol06} Solanki, S.K., Inhester, B., \& Sch\"ussler, M., 2006, Rep.Prog.Phys., 69, 563-668
\bibitem[Sorriso-Valvo et al(2001)]{sor01} Sorriso-Valvo, L., Carbone, V., Giuliani, P., Veltri, P., Bruno, R., Antoni, V., \& Martines, E., 2001, Planetary and Space Science, 49, 12, 1193
\bibitem[Soward et al Eds.(2002)]{sow02} Soward, A.M., Jones, C.A., Hughes, D.W., \& Weiss, N.o., (Eds.), 2002, Fluid dynamics and dynamos in astrophysics and geophysics (Florida: CRC Press)
\bibitem[Subramanian et al.(2006)]{sub06} Subramanian, K., Shukurov, A., \& Haugen, N.E.L., 2006, \mnras, 366, 1437
\bibitem[Subramanian (2010)] {sub10} Subramanian, K., 2010, Astron.Nachr./AN, 331, 1, 110
\bibitem[Stix(2002)]{sti02} Stix, M., 2002, The Sun, an Introduction (Berlin, New York: Springer)
\bibitem[Taylor(1974)]{tay74} Taylor, J.B., 1974, Phys. Rev.Lett, 33, 19, 1139
\bibitem[van Weeren et al.(2009)]{van09} van Weeren, R.J., R\"ottgering, J.A., Bagchi, J., Raychaudhury, S., Intema, H.T., Miniati, F., En{\ss}lin,T.A., Markevitch, M., \& Erben, T.,  2009, A\&A, 506, 1083
\bibitem[ Waleffe(1992)]{wa192} Waleffe, F., 1992, Physics of Fluids A, 4(2), 350
\bibitem[Weiss(2002)]{wei02} Weiss, N., 2002, A\&G, 43, 3.9-3.14
\bibitem[Widrow(2002)]{wi02} Widrow, L.M., 2002, Reviews of Modern Physics, 74, 775
\bibitem[Yoshimatsu et al(2013)]{yos12} Yoshimatsu, K., Okamoto, N., Kawahara, Y., Schneider, K.,  \& Farge, M., 2013, Geophysical \& Astrophysical Fluid Dynamics, 107, 12, 73 DOI:10.1080/03091929.2012.654790
\bibitem[Zhou et al.(2004)]{zh04} Zhou, Y., Matthaeus, W.M., \& Dmitruk, P., 2004, Reviews of Modern Physics, 76, 1015
\end{thebibliography}
\end{document}